\documentclass[conference]{IEEEtran}
\IEEEoverridecommandlockouts

\usepackage{cite}
\usepackage{amsmath,amssymb,amsfonts}
\usepackage{algorithmic}
\usepackage{graphicx}
\usepackage{textcomp}
\usepackage{xcolor}
\def\BibTeX{{\rm B\kern-.05em{\sc i\kern-.025em b}\kern-.08em
    T\kern-.1667em\lower.7ex\hbox{E}\kern-.125emX}}

\usepackage{microtype}
\usepackage{graphicx}
\usepackage{booktabs} %
\usepackage{multirow}
\usepackage{multicol}
\usepackage{tablefootnote}

\usepackage[hidelinks]{hyperref}

\usepackage{amsmath}
\usepackage{amssymb}
\usepackage{mathtools}
\usepackage{amsthm}
\usepackage{caption}
\usepackage{subcaption}
\usepackage{booktabs}
\usepackage[flushleft]{threeparttable}
\usepackage{MnSymbol}
\usepackage{lipsum}

\usepackage{bigstrut}
\usepackage{colortbl}
\usepackage{wrapfig}

\usepackage{enumitem}
\setlist[itemize]{noitemsep}
\setlist[enumerate]{noitemsep}

\usepackage[ruled, vlined, linesnumbered]{algorithm2e} %

\usepackage[capitalize,noabbrev]{cleveref}

\usepackage{makecell}

\usepackage{bm}
\newcommand\bolden[1]{{\boldmath\bfseries#1}}

\usepackage{contour}
\usepackage[normalem]{ulem}
\contourlength{0.8pt}

\newcommand{\myuline}[1]{%
  \uline{\phantom{#1}}%
  \llap{\contour{white}{#1}}%
}

\newcommand{\smallsection}[1]{\vspace{0.5mm}{\noindent {\bolden{\myuline{#1}}}}}

\usepackage{xspace}

\def\mydefbb#1{\expandafter\def\csname bb#1\endcsname{\ensuremath{\mathbb{#1}}}}
\def\mydefallbb#1{\ifx#1\mydefallbb\else\mydefbb#1\expandafter\mydefallbb\fi}
\mydefallbb ABCDEFGHIJKLMNOPQRSTUVWXYZ\mydefallbb

\def\mydefcal#1{\expandafter\def\csname cal#1\endcsname{\ensuremath{\mathcal{#1}}}}
\def\mydefallcal#1{\ifx#1\mydefallcal\else\mydefcal#1\expandafter\mydefallcal\fi}
\mydefallcal ABCDEFGHIJKLMNOPQRSTUVWXYZ\mydefallcal

\newcommand{\setbr}[1]{\{#1\}}
\newcommand{\mtsetbr}[1]{\left\{\mkern-6mu\left\{#1\right\}\mkern-6mu\right\}}

\newcommand{\ours}{\textsc{AnchorRadar}\xspace}

\definecolor{lucky}{RGB}{220, 20, 60}
\definecolor{kjgreen}{RGB}{77, 175, 74}

\newcommand\revisekdd[1]{{#1}}
\newcommand\revisekddcolor{\color{black}}

\makeatletter
\newtheorem*{rep@theorem}{\rep@title}
\newcommand{\newreptheorem}[2]{%
\newenvironment{rep#1}[1]{%
 \def\rep@title{#2 \ref{##1}}%
 \begin{rep@theorem}}%
 {\end{rep@theorem}}}
\makeatother

\theoremstyle{plain}

\newreptheorem{theorem}{Theorem}

\newreptheorem{proposition}{Proposition}

\newreptheorem{property}{Property}

\newreptheorem{lemma}{Lemma}

\newreptheorem{corollary}{Corollary}

\theoremstyle{definition}
\newtheorem{definition}{Definition}

\newtheorem{observation}{Observation}

\theoremstyle{remark}

\newtheorem{problem}{Problem}

\usepackage{enumitem}
\usepackage{multirow,multicol}
\usepackage{bigstrut}
\usepackage{adjustbox}

\SetKwComment{Comment}{$\triangleright$\ }{}

\SetCommentSty{mycommfont}
\SetAlFnt{\small}

\usepackage{placeins}

\setlist{nolistsep}

\let\citep\cite
\let\citet\cite
\begin{document}

\title{Identifying Group Anchors in Real-World\\Group Interactions Under Label Scarcity}

\author{
Fanchen Bu\textsuperscript{1},
Geon Lee\textsuperscript{2},
Minyoung Choe\textsuperscript{2}, and
Kijung Shin\textsuperscript{1,2}\\
\textsuperscript{1}School of Electrical Engineering and
\textsuperscript{2}Kim Jaechul Graduate School of AI, KAIST, Republic of Korea\\
\{boqvezen97, geonlee0325, minyoung.choe, kijungs\}@kaist.ac.kr
}
\maketitle

\nocite{onlineSuppl}
\begin{abstract}
Group interactions occur in various real-world contexts, e.g., co-authorship, email communication, and online Q\&A.
\textit{In each group}, there is often a particularly significant member, around whom the group is formed.
Examples include the first or last author of a paper, the sender of an email, and the questioner in a Q\&A session.
In this work, we discuss the existence of such individuals in real-world group interactions.
We call such individuals group \textit{anchors} and study the problem of identifying them.
First, we introduce the concept of group anchors and the identification problem.
Then, we discuss our observations on group anchors in real-world group interactions.
Based on our observations, we develop \ours, a fast and effective method for group anchor identification under realistic settings with label scarcity, i.e., when only a few groups have known anchors.
\ours is a semi-supervised method using information from groups both with and without known group anchors.
Finally, through extensive experiments on thirteen real-world datasets, we demonstrate the empirical superiority of \ours over various baselines w.r.t. accuracy and efficiency.
In most cases, \ours achieves higher accuracy in group anchor identification than all the baselines, while using 10.2$\times$ less training time than the fastest baseline and 43.6$\times$ fewer learnable parameters than the most lightweight baseline on average.
\end{abstract}

\begin{IEEEkeywords}
Group interactions, Hypergraphs, Label scarcity, Important node identification
\end{IEEEkeywords}

\section{Introduction}\label{sec:intro}

Group interactions are ubiquitous in the real world.
For example, scholars collaborate and write a paper together~\citep{lung2018hypergraph}, people participate in the same email communication~\citep{larock2023encapsulation}, users participate in the same online Q\&A session~\citep{antelmi2023age}, and actors act in the same movie~\citep{yadati2024heal}.
In many scenarios, in each group, a particular member plays a specifically important role, around whom the group is formed~\citep{hinds2000choosing}.
For example, in an academic collaboration, the first or last author of a paper is often the one who connects all the authors~\citep{osborne2019authorship}.
In an email correspondence, the sender of an email is the one who initializes the conversation to include the recipients~\citep{pallen1995guide}.
In an online Q\&A session, the questioner starts the session for others to join~\citep{stackexchangeTourStack}.
In a movie cast, the ensemble is often gathered around the main actor~\citep{colliderMoviesStarring}.

In this work, we discuss the existence of such important individuals and call them group \textit{anchors}, in the sense that each brings together the members of a group.
As widely observed across different domains (see \cref{tab:single_node_label}), there is often a single anchor in each group (see Section~\ref{sec:experiments:acc} for discussions and results with multiple anchors in each group).
Notably, the concept of group anchors is defined by the real-world meaning of groups and the roles of members, which is not necessarily aligned with (sometimes even opposite to) structural importance.
For example, the anchor (i.e., questioner) of an online Q\&A session is often a new user with low structural importance.
Finding group anchors can help us better understand and analyze real-world group interactions, and has the following important further real-world applications:
\begin{itemize}[leftmargin=*]
    \item \textbf{Group-interaction prediction.} 
    Group anchors often initiate or determine the formation of group interactions. By identifying anchors, we can better predict how groups form and evolve over time~\citep{benson2018simplicial} for, e.g., academic collaboration recommendation~\citep{liu2018context} and social group recommendation~\citep{qin2018dynamic}.
    \item \textbf{Engagement management.} Group anchors play significant roles in group interactions and need particular attention.
    In social networks, identifying anchors provides insights into the cohesion, longevity, and activity levels of social groups, helping administrators better maintain group engagement~\citep{malliaros2013stay}.
    \item \textbf{Targeted marketing.} Group anchors, being central or highly influential in their groups, can be critical targets for marketing. By focusing on group anchors, marketers can effectively influence whole groups through just a few individuals~\citep{chandra2022personalization}.
\end{itemize}
To the best of our knowledge, we are the first to discuss the existence of group anchors and study the identification problem.

\begingroup
\setlength{\tabcolsep}{4pt}
\begin{table}[t!]
    \centering
    \caption{Examples of group anchors in real-world group interactions.}
    \label{tab:single_node_label}
\begin{adjustbox}{max width=\linewidth}
\begin{tabular}{c| c c}
\hline
\textbf{Domain} & \textbf{Members in Each Group} & \textbf{Anchor in Each Group} \bigstrut\\
\hline
Co-authorship & Authors of a paper & First/last author \bigstrut[t]\\
Online Q\&A & Users in a session & Questioner \\
Email & People in a correspondence & Sender \\
Message & People involved a message & Sender \\
Retweet & People involved a retweet & Retweeted \\
Movie & Actors in a movie & Leading actor \bigstrut[b]\\
\hline
\end{tabular}%
\end{adjustbox}
\vspace{-1mm}
\end{table}
\endgroup

First, we introduce the concept of group anchors and the identification problem.
For mathematical rigor, we formulate the problem as an optimization problem on hypergraphs.
Hypergraphs allow an arbitrary number of nodes in each hyperedge and precisely model group interactions~\citep{benson2018sequences}.
Each hyperedge represents an interacting group, with its constituent nodes as the group members.
The anchor of each group is a node in the corresponding hyperedge.

Then, we discuss our observations on real-world group anchors.
We focus on the scenarios under label scarcity with limited training data (i.e., only a few groups have known anchors), which is common in the real world ~\citep{zhu2022introduction,zhu2005semi,yang2022semi}.
What can we learn with limited known group anchors?
\textbf{\textit{Our first observation}} is that topological features provide considerable information on group anchors, but they alone are insufficient.
Specifically, a simple heuristic based on node degrees 
performs comparably to sophisticated deep-learning methods sometimes, but it does not fully capture the underlying mechanisms of group anchors, leaving a clear gap compared to the strongest baseline.
Also, we would like to re-emphasize that, in some scenarios, structurally insignificant nodes are more likely to be the group anchors (e.g., questioners in online Q\&A).
\textbf{\textit{Our second observation}} is that whether an individual is the anchor or not is overall stable, yet still contextual, across different groups.
It is ``stable'' in that if an individual is (not) the anchor in one group, they are likely (not) the anchor in other groups.
It is ``contextual'' in that it is still possible for an individual to be the anchor in some groups but not in others.
We further observe that it is essential to consider \textit{local competition} in each group on top of \textit{global stability} to better capture the mechanism of group anchors.
Specifically, group anchors are well explained by
\textit{(i)} each node has its ``anchor strength'' indicating the overall likelihood that the node is the anchor across different groups, and
\textit{(ii)} the nodes compete locally in each group, and the one with the highest anchor strength is likely the anchor.

Intuitively and explicitly based on our observations, we propose \ours, a novel method for identifying group anchors.
Our method \ours has two stages.
\textbf{\textit{In the first stage}}, motivated by our first observation, we train a model to utilize the information in topological features to match the known group anchors.
After the first-stage model is trained, its predictions are used as ``references'' for the second stage.
\textbf{\textit{In the second stage}}, motivated by our second observation, we associate each node with a single scalar (i.e., ``anchor strength''), and we train the scalars to match the known group anchors and the ``references'' from the first stage.
After the second-stage strengths are trained, the node with the highest strength in each hyperedge is predicted as the anchor.

Instead of using sophisticated architectures such as deep neural networks, we aim to make \ours efficient and lightweight by using only simple architectures with few parameters.
This two-stage training balances fidelity to the given topology in the first stage and flexibility in the second stage.
Notably, \ours is semi-supervised, using information from groups both with and without known group anchors, e.g., the topological features.
In summary, the proposed method \ours is
\textbf{\textit{(i) intuitive}}, with algorithmic designs directly based on observations on real-world datasets, and
\textbf{\textit{(ii) efficient}}, using few parameters and demanding less computational resources, e.g., training time.
The empirical superiority of \ours is validated via extensive experiments on thirteen real-world hypergraph datasets.

To conclude, our contributions are four-fold:
\begin{itemize}[leftmargin=*]
    \item \textbf{New concept and new problem (\cref{sec:problem_state}).} To the best of our knowledge, we are the first to discuss the existence of group anchors, and study the problem of identifying them in real-world group interactions under realistic settings with label scarcity.
    This problem is of theoretical interest with various further practical applications.
    \item \textbf{Observations (\cref{sec:observations}).} Based on the realistic scenarios with limited known group anchors, we discuss several observations on group anchors in real-world group interactions.
    \item \textbf{Method (\cref{sec:method}).} We propose \ours, a novel method for group anchor identification, with intuitive and lightweight mechanisms motivated by our observations.
    \item \textbf{Experiments (\cref{sec:experiments}).} Via experiments on thirteen real-world hypergraphs, we validate the empirical superiority of our method \ours w.r.t. accuracy and efficiency.
\end{itemize}

\smallsection{Reproducibility.} 
The appendix, code, and datasets are available online in~\citep{onlineSuppl} (\url{https://github.com/bokveizen/anchor_radar}).

\vspace{-3mm}
\section{Preliminaries}\label{sec:prelim}

\smallsection{Basic notations.}
We use $\bbN \coloneqq \setbr{1, 2, 3, \ldots}$ to denote the set of natural numbers, and $[n] \coloneqq \setbr{1, 2, 3, \ldots, n}$ to denote the set of all natural numbers at most $n$.
We use $\mtsetbr{\cdot}$ explicitly for \textit{multisets} allowing duplicated elements and $\setbr{\cdot}$ for \textit{sets} consisting of unique elements.

\smallsection{Hypergraphs.}
We use hypergraphs to represent group interactions.
Each hyperedge corresponds to a group where the nodes represent the members.
Formally, a \textit{hypergraph} $H = (V, E)$ is defined by its \textit{topology} consisting of a \textit{node} set $V$ and a \textit{hyperedge} multiset $E$, where each hyperedge $e \in E$ is a subset of $V$, i.e., $e \subseteq V$.
We use $E^* = \setbr{e: e \in E}$ to denote the set of unique hyperedges after removing repetitions.
The \textit{degree} $d_v$ of a node $v$ is the number of hyperedges containing $v$, i.e., $d_v = |\mtsetbr{e \in E: v \in e}|$.
We assume no given node or edge attributes (i.e., features), which is the case for the real-world datasets used in our experiments, and also many other datasets~\citep{lee2024villain}.
Yet, attributes can be easily incorporated into our method when given. See Section~\ref{sec:conclusion} for more discussions.
See Appendix~\ref{appx:discussions} for discussions on general hypergraphs, e.g., directed and heterogeneous hypergraphs.

\begingroup
\setlength{\tabcolsep}{5pt}
\begin{table*}[t!]
    \centering
    \caption{Basic statistics of the real-world group-interaction (hypergraph) datasets used in our experiments.}
    \label{tab:dataset_summary}
\begin{adjustbox}{max width=\textwidth}
\begin{tabular}{ccc|rrr|rrr}
\hline
\textbf{Domain} & \textbf{Dataset} & \textbf{Abbrev.} & $\mathbf{|V|}$ & $\mathbf{|E|}$ & $\mathbf{|E^*|}$ & \textbf{Min.} $\mathbf{|e|}$ & \textbf{Max.} $\mathbf{|e|}$ & \textbf{Avg.} $\mathbf{|e|}$ \bigstrut\\
\hline
\multirow{3}[2]{*}{\makecell{Co-authorship \\ ($D_{\text{co}}$)}} & AMinerAuthor~\citep{tang2008arnetminer,choe2023classification}  & \texttt{coAA} & 1,712,433 & 2,037,605 & 1,454,250 & 1     & 115   & 2.55  \bigstrut[t]\\
\multicolumn{1}{c}{} & DBLP~\citep{team2019dblp,choe2023classification} & \texttt{coDB} &  108,476 & 91,260 & 81,601 & 2     & 36    & 3.52 \\
\multicolumn{1}{c}{} & ScopusMultilayer~\citep{chodrow2020annotated,kivela2014multilayer,bianconi2018multilayer,boccaletti2014structure} & \texttt{coSM} & 1,673 & 937   & 842   & 1     & 27    & 3.09 \bigstrut[b]\\
\hline
\multirow{4}[2]{*}{\makecell{Online Q\&A \\ ($D_{\text{qa}}$)}} & StackOverflowBiology~\citep{choe2023classification} & \texttt{qaBI} &  15,418 & 26,290 & 23,242 & 1     & 12    & 2.08  \bigstrut[t]\\
\multicolumn{1}{c}{} & StackOverflowPhysics~\citep{choe2023classification} & \texttt{qaPH} & 80,434 & 194,575 & 169,274 & 1     & 40    & 2.38 \\
\multicolumn{1}{c}{} & MathOverflow~\citep{chodrow2020annotated} & \texttt{qaMA} & 410   & 154   & 154   & 2     & 57    & 4.27  \\
\multicolumn{1}{c}{} & StackOverflow~\citep{chodrow2020annotated} & \texttt{qaST} &  22,131 & 4,716 & 4,713 & 1     & 59    & 5.79 \bigstrut[b]\\
\hline
\multirow{3}[2]{*}{\makecell{Email \\ ($D_{\text{em}}$)}} & EmailEnron~\citep{choe2023classification} & \texttt{emEN} & 21,251 & 101,124 & 34,916 & 2     & 883   & 11.53 \bigstrut[t]\\
\multicolumn{1}{c}{} & EmailEu~\citep{choe2023classification,paranjape2017motifs} & \texttt{emEU} & 986   & 209,508 & 24,520 & 2     & 40    & 2.56 \\
\multicolumn{1}{c}{} & Enron~\citep{chodrow2020annotated} & \texttt{emER} & 110   & 9,603 & 1,169 & 2     & 29    & 2.47  \bigstrut[b]\\
\hline
\multirow{2}[2]{*}{Social network ($D_{\text{so}}$)} & Message~\citep{chodrow2020annotated} & \texttt{soME} & 26,059 & 34,577 & 22,700 & 2     & 14    & 2.58 \bigstrut[t]\\
\multicolumn{1}{c}{} & Retweet~\citep{chodrow2020annotated} & \texttt{soRE} & 30,073 & 88,148 & 49,828 & 2     & 2     & 2.00   \bigstrut[b]\\
\hline
Movie cast ($D_{\text{mo}}$)  & MovieLens~\citep{chodrow2020annotated,harper2015movielens} & \texttt{moML} & 73,155 & 43,058 & 42,497 & 1     & 5     & 4.70  \bigstrut\\
\hline
\multicolumn{9}{l}{\small *Data source: \url{https://github.com/young917/EdgeDependentNodeLabel}~\citep{choe2023classification} and \url{https://andrewmellor.co.uk/data/}~\citep{chodrow2020annotated}.} \\
\end{tabular}%
\end{adjustbox}        
\end{table*}
\begingroup

\section{Related Work}\label{sec:rel_wk}

\smallsection{Important nodes in hypergraphs.}
Typically, \textit{global} (i.e., for the whole hypergraph) and \textit{structural} (i.e., based on hypergraph topology) centrality measures are used to identify important nodes, where nodes with high centrality scores are considered important.
Some examples of node centrality measures include variants of core numbers~\citep{arafat2023neighborhood,ramadan2004hypergraph,limnios2021hcore,lee2023k,kim2023exploring,bu2023hypercore} and various spectral centrality measures~\citep{tudisco2021node,kovalenko2022vector,benson2019three}.
Specifically, the core-periphery structure in hypergraphs has been considered~\citep{tudisco2023core,papachristou2022core}, where the set of ``core nodes'' is typically a small set of nodes intersecting with most (or even all~\citep{amburg2021planted}) hyperedges.
In summary, most existing works identify important nodes based on their \textit{global} \textit{structural} properties, while by the idea of group anchors, we consider the \textit{conceptual} (i.e., based on the real-world meaning of the group interactions) importance of nodes \textit{locally} in each hyperedge (group).

\smallsection{Node classification in hypergraphs.}
Node classification is a common task on 
hypergraphs~\citep{feng2019hypergraph}, where one aims to assign labels to nodes.
General node classification considers node labels for a node the same across different hyperedges, i.e., the node labels are edge-independent. Therefore, methods for general node classification cannot be directly applied to identifying group anchors, which are edge-dependent.\footnote{See Appendix~\ref{appx:general_node_cls} for more discussions on general node classification.}
Indeed, as noted in \citet{chodrow2020annotated}, the same entity (i.e., node) may have different roles (i.e., labels) in different groups (i.e., hyperedges), which motivated prior works on the task of edge-dependent node classification (ENC) in hypergraphs~\citep{choe2023classification, zheng2024co}.
{WHATsNet~\citep{choe2023classification} tackles ENC by generating node and hyperedge embeddings using SetTransformer~\citep{lee2019set} with positional encoding, and aggregating them to obtain the embeddings of node-hyperedge pairs.
CoNHD~\citep{zheng2024co} uses co-representation hypergraph diffusion to directly learn node-hyperedge pair embeddings by considering interactions both between hyperedges and between nodes.}
Both WHATsNet and CoNHD are built on deep hypergraph neural network architectures.
Mathematically, identifying group anchors can be seen as a special case of ENC, where the classification is binary, and there is one positive sample (i.e., the anchor) in each hyperedge.
Indeed, we apply methods for ENC to group anchor identification with proper modifications as baseline methods.
However, the problem of group anchor identification has unique values and applications (see \cref{sec:intro}).
Also, as shown in our experiments (see \cref{sec:experiments}), even though using much more sophisticated mechanisms, existing methods for ENC are outperformed by our method w.r.t. both accuracy and efficiency.
Moreover, we consider realistic settings with label scarcity, while existing works on edge-dependent node classification~\citep{choe2023classification,zheng2024co} consider settings where the node labels in 60\% of the hyperedges (groups) are known.

\section{Concepts and Problem}\label{sec:problem_state}

We shall introduce the concept of \textit{group anchors} and the problem of \textit{group anchor identification}. 
To the best of our knowledge, we are the first to formulate and study them.

\smallsection{Domains and group roles.}
We consider real-world hypergraphs (group interactions) in different \textit{domains} $\mathcal{D} = \setbr{D_{\text{co}}, D_{\text{qa}}, D_{\text{em}}, D_{\text{so}},  D_{\text{mo}}, \ldots}$, where 
$D_{\text{co}}$ is the co-authorship domain, 
$D_{\text{qa}}$ is the online Q\&A domain, 
$D_{\text{em}}$ is the email domain, and so on (see \cref{tab:dataset_summary}).
For each domain $D \in \mathcal{D}$, let $\mathcal{R}(D)$ be the set of \textit{group roles} (i.e., the roles members can take in each group) in that domain, e.g., the group roles in the co-authorship domain are $\mathcal{R}(D_{\text{co}}) = \setbr{\texttt{firstAuthor}, \texttt{middleAuthor}, \texttt{lastAuthor}}$.
The domain $D(H) \in \mathcal{D}$ indicates the real-world meaning of the hyperedges (groups) in $H$,
and in each hyperedge (group) $e \in E$, each node $v \in e$ has a group role $r(v; e) \in \mathcal{R}(D(H))$.

\smallsection{Group anchors.}
The \textit{group-anchor function} $f_{\text{anchor}}$ maps each domain to the group role of anchors in that domain.
Given a domain $D \in \mathcal{D}$, $f_{\text{anchor}}(D) \in \mathcal{R}(D)$ is the group role of group anchors in domain $D$.
For example, $f_{\text{anchor}}(D_{\text{qa}}) = \texttt{questioner}$, and 
$f_{\text{anchor}}(D_{\text{em}}) = \texttt{sender}$.
See Appendix~\ref{appx:roles_and_seed_members} for more details.
Given a hypergraph $H = (V,E)$ in domain $D = D(H)$, the set of group anchors in each hyperedge $e \in E$ is $A(e;H) = \setbr{v \in e : r(v; e) = f_{\text{sm}}(D)}$.
As mentioned in Section~\ref{sec:intro}, we focus on the scenario with a single anchor in each group (i.e., $|A(e;H)| \equiv 1$),
as widely observed across different domains (see \cref{tab:single_node_label}).
See Section~\ref{sec:experiments:acc} for discussions and results considering multiple group anchors.

\color{black}

\smallsection{Group anchor identification.}
We formulate group anchor identification as an optimization problem on hypergraphs.
\begin{problem}[Group anchor identification]\label{prob:seed_identification}
 ~\\[-1em]
\begin{itemize}[leftmargin=*]    
    \item \textbf{Given:} A hypergraph $H = (V, E)$ with known group anchors in a hyperedge subset $E' \subseteq E$ (i.e., $A(e), \forall e \in E'$);
    \item \textbf{To Predict:} The unknown group anchors in the remaining hyperedges $E \setminus E'$ (i.e., $A(e), \forall e \in E \setminus E'$);
    \item \textbf{to Maximize:} The accuracy of the predicted group anchors.
\end{itemize}
\end{problem}

As discussed in \cref{sec:prelim}, we assume no node or edge attributes (i.e., features) are given, which is true for the real-world datasets used in our experiments, and also many other datasets~\citep{lee2024villain}. Yet, attributes can be easily incorporated into our method if they are given. See Section~\ref{sec:conclusion} for more discussions.

\section{Observations on Real-World Datasets}\label{sec:observations}

We shall describe several observations we have on the group anchors in real-world group interactions.

Note that we assume realistic settings with label scarcity~\citep{zhu2022introduction,zhu2005semi,yang2022semi}, where the anchors in only a low proportion of groups are known.
Specifically, we use 7.5\% training data and 2.5\% validation data in our main experiments (see Section~\ref{sec:experiments} for more details; \revisekdd{see also Appendix~\ref{appx:diff_training_ratios} for additional results with different training ratios}).
Hence, when we discuss observations, we also assume 7.5\% groups with known anchors, and show that our observations are well established even with only such a low proportion of known group anchors.

For each observation, we shall first provide the statement, then real-world intuitions behind the observation, and finally, empirical evidence from the real-world hypergraphs we use.

\smallsection{Datasets.}
For ease of discussion on our observations, we shall first introduce the real-world group-interaction (hypergraph) datasets used in this work, with some details deferred to \cref{sec:experiments} on experiments.
We collect datasets from existing works~\citep{choe2023classification,chodrow2020annotated}.
Specifically, we use thirteen datasets from five different domains:
\begin{itemize}[leftmargin=*]
    \item \textbf{Co-authorship (\texttt{co}) datasets}:
    Each node represents a scholar, and each hyperedge includes the coauthors of the same publication.
    In each hyperedge, the group anchor is the author who connects and gathers the other authors.
    Arguably, either the first or the last author can be the anchor~\citep{osborne2019authorship}. Therefore, in this work, we consider both possible cases.
    \item \textbf{Online Q\&A (\texttt{qa}) datasets}:
    Each node represents a user, and each hyperedge includes those involved in a Q\&A session.
    In each hyperedge, the group anchor is the questioner, who opens the session so that the other users can join it.
    \item \textbf{Email (\texttt{em}) datasets}:
    Each node represents a person, and each hyperedge includes those involved in the same email correspondence.
    In each hyperedge, the anchor is the sender, who initially includes all the recipients in the correspondence.
    \item \textbf{Social network (\texttt{so}) datasets}:
    Each node represents a user, and each hyperedge includes those involved in the same social interaction (message or retweet).
    In each hyperedge, the group anchor is the person initiating the interaction, i.e., the sender of the message, who initially chooses to include all the recipients, or the retweeted user, who uploads the original post so that the others can retweet it.    
    \item \textbf{Movie cast (\texttt{mo}) dataset}:
    Each node represents an actor, and each hyperedge includes those acting in the same movie.
    In each hyperedge, the group anchor is the leading actor, typically around whom the whole ensemble is built~\citep{colliderMoviesStarring}.
\end{itemize}
See \cref{tab:single_node_label} for a summary of the datasets.
See Table~\ref{tab:dataset_summary} for the basic statistics of the datasets.
More details on data processing are deferred to Section~\ref{sec:exp_settings} when discussing experiments.

\begingroup
\setlength{\tabcolsep}{3pt}
\begin{table}[t!]
    \centering   
    \caption{\uline{Observation 1: Topological features are informative about group anchors, yet are limited.} Predicting the node with the highest or lowest degree in each hyperedge as the anchor achieves considerable, yet not outstanding, accuracy (\%).
    In each setting, the best performance is highlighted in bold, while the second-best is underlined.}
    \label{tab:degree_perf}
\begin{adjustbox}{max width=\linewidth}
\begin{tabular}{cc|ccccc}
\hline
\multicolumn{2}{c|}{\textbf{Dataset}} & {\small Degree} & {\small WHATsNet} & {\small CoNHD-U} & {\small CoNHD-I} & {\small Random} \bigstrut \\
\hline
\multirow{2}[0]{*}{\texttt{coAA}} & (first) & 44.4  & \textbf{45.2} & 42.7  & \uline{44.5} & 37.1 \bigstrut[t] \\
      & (last) & \textbf{46.3} & \uline{45.8} & 42.2  & 44.7  & 37.1 \\
\multirow{2}[0]{*}{\texttt{coDB}} & (first) & \textbf{42.5} & \uline{42.5} & 41.2  & 41.4  & 32.8 \\
      & (last) & \textbf{45.5} & \uline{45.4} & 42.7  & 43.5  & 32.8 \\
\multirow{2}[0]{*}{\texttt{coSM}} & (first) & 32.7  & \textbf{34.3} & 31.4  & 29.8  & \uline{33.7} \\
      & (last) & 37.7  & \textbf{39.8} & 37.9  & \uline{39.4} & 33.7 \\
\multicolumn{2}{c|}{\texttt{qaBI}} & 76.2  & \textbf{85.6} & 78.7  & \uline{79.3} & 44.3 \\
\multicolumn{2}{c|}{\texttt{qaPH}} & 77.2  & \textbf{88.1} & 76.0  & \uline{77.3} & 41.1 \\
\multicolumn{2}{c|}{\texttt{qaMA}} & \textbf{38.7} & \uline{35.8} & 29.0  & 29.8  & 32.4 \\
\multicolumn{2}{c|}{\texttt{qaST}} & \uline{30.9} & \textbf{31.2} & 25.4  & 26.6  & 24.2 \\
\multicolumn{2}{c|}{\texttt{emEN}} & 22.0  & \textbf{50.8} & 44.0  & \uline{45.1} & 18.8 \\
\multicolumn{2}{c|}{\texttt{emEU}} & 49.0  & \textbf{51.0} & 52.8  & \uline{52.4} & 45.8 \\
\multicolumn{2}{c|}{\texttt{emER}} & \uline{66.2} & \textbf{66.6} & 65.3  & 64.6  & 44.9 \\
\multicolumn{2}{c|}{\texttt{soME}} & 65.0  & \textbf{75.5} & 74.3  & \uline{74.6} & 42.9 \\
\multicolumn{2}{c|}{\texttt{soRE}} & 84.3  & \uline{97.4} & 96.8  & \textbf{97.5} & 50.0 \\
\multicolumn{2}{c|}{\texttt{moML}} & 41.8  & 41.4  & \uline{42.4} & \textbf{42.7} & 21.3 \bigstrut[b]\\
\hline
\multicolumn{2}{c|}{Avg. Acc.} & 50.0  & \textbf{54.8} & 51.4  & \uline{52.1} & 35.8 \bigstrut[t]\\
\multicolumn{2}{c|}{Avg. Rank} & 2.75  & \textbf{1.63} & 3.38  & \uline{2.56} & 4.69 \bigstrut[b]\\
\hline
\end{tabular}%
\end{adjustbox}
\end{table}
\endgroup

\subsection{Observation 1: Informative Yet Limited Topological Features (\cref{tab:degree_perf})}\label{sec:intuition:topological_feat}

The most immediate information we have is the topology (i.e., nodes and hyperedges; see \cref{prob:seed_identification}), especially under label scarcity.
From topology, we can derive topological features. 
However, are topological features actually helpful in identifying anchors in real-world group interactions?
Our first observation gives a partially positive answer to this question.

\vspace{-1mm}
\begin{observation}\label{intu:topological_feat}
    In real-world group interactions, topological features are informative about group anchors.
    That is, one can achieve considerable, yet not outstanding, accuracy in identifying group anchors based on only topological features.
\end{observation}
\vspace{-1mm}

\smallsection{Intuition.}
Let us first provide concrete intuition in different real-world domains behind the observation.
Specifically, let us consider one of the simplest topological features, node degrees (see \cref{sec:prelim}).
In co-authorship datasets, nodes with high degrees are likely senior scholars, who are also likely the last authors, but early-career scholars can also sometimes be the last authors (e.g., a new professor with her students).
In online Q\&A datasets, nodes with low degrees are likely new users in the forum, who are also likely to ask questions, but experienced users can also sometimes ask questions.
In movie cast datasets, nodes with high degrees are likely famous actors, who are also likely the leading actors, but novice actors (e.g., ``rising stars'') can also sometimes be the leading actors.

\smallsection{Evidence.}
Below, we provide more detailed evidence of our observation.
We compare the performance of a simple heuristic based on only node degrees, with state-of-the-art deep-learning methods (WHATsNet~\citep{choe2023classification} and CoNHD~\citep{zheng2024co}) for edge-dependent node classification (ENC; see \cref{sec:rel_wk}), which we adapt for group anchor identification.
For each dataset, between the highest and lowest degree, the heuristic chooses the better one based on their accuracy on the 7.5\% training data with known group anchors.
Notably, WHATsNet~\citep{choe2023classification} and CoNHD~\citep{zheng2024co} also use topological features (including node degrees).
The detailed experimental settings will be introduced later in \cref{sec:exp_settings}.
As shown in \cref{tab:degree_perf}, in many real-world hypergraphs, simply predicting the node with the highest or lowest degree in each hyperedge has considerable accuracy.
Yet, a clear gap remains between such a purely topological feature-based heuristic and the strongest baseline WHATsNet.
We also include the accuracy of random guesses as a reference.

\begingroup
\setlength{\tabcolsep}{6pt}
\begin{table}[t!]
    \centering
    \caption{\uline{Observation 2: Whether a node is the group anchor or not is stable, yet not fully homogeneous, across different hyperedges.} The anchor purity in real-world datasets is significantly higher than in randomized ones in most cases, indicated by the small $p$-values obtained from one-sided $t$-tests, but the purity is not near 100\% in most cases.
    \revisekdd{We report statistics on only 7.5\% hyperedges to match our experimental settings with the same amount of training data. See Table~\ref{tab:obs_2_all} in Appendix~\ref{appx:obs_2_all} for statistics on all hyperedges, where the statistical significance of the observation is even stronger.}}
    \label{tab:label_purity}
\begin{adjustbox}{max width=\linewidth}
\begin{tabular}{cc|cc|c}
\hline
\multicolumn{2}{c|}{\textbf{Dataset}} & Real-world & Random & $p$-value \bigstrut\\
\hline
\multirow{2}[1]{*}{\texttt{coAA}} & (first) & 0.7420{\scriptsize $\pm$0.3706} & 0.5762{\scriptsize $\pm$0.4012} & $<$0.0001 \bigstrut[t]\\
      & (last) & 0.7375{\scriptsize $\pm$0.3662} & 0.5758{\scriptsize $\pm$0.4012} & $<$0.0001 \\
\multirow{2}[0]{*}{\texttt{coDB}} & (first) & 0.7708{\scriptsize $\pm$0.3786} & 0.5873{\scriptsize $\pm$0.4242} & $<$0.0001 \\
      & (last) & 0.7490{\scriptsize $\pm$0.3801} & 0.5977{\scriptsize $\pm$0.4209} & $<$0.0001 \\
\multirow{2}[0]{*}{\texttt{coSM}} & (first) & 0.7821{\scriptsize $\pm$0.3777} & 0.5103{\scriptsize $\pm$0.4728} & 0.0146 \\
      & (last) & 0.8872{\scriptsize $\pm$0.2335} & 0.5577{\scriptsize $\pm$0.4598} & 0.0012 \\
\multicolumn{2}{c|}{\texttt{qaBI}} & 0.8196{\scriptsize $\pm$0.3248} & 0.5323{\scriptsize $\pm$0.3692} & $<$0.0001 \\
\multicolumn{2}{c|}{\texttt{qaPH}} & 0.8146{\scriptsize $\pm$0.3239} & 0.5375{\scriptsize $\pm$0.3700} & $<$0.0001 \\
\multicolumn{2}{c|}{\texttt{qaMA}} & 0.8750{\scriptsize $\pm$0.3307} & 0.6250{\scriptsize $\pm$0.4841} & 0.1391 \\
\multicolumn{2}{c|}{\texttt{qaST}} & 0.9051{\scriptsize $\pm$0.2696} & 0.7551{\scriptsize $\pm$0.3987} & 0.0141 \\
\multicolumn{2}{c|}{\texttt{emEN}} & 0.9430{\scriptsize $\pm$0.1700} & 0.8551{\scriptsize $\pm$0.2408} & $<$0.0001 \\
\multicolumn{2}{c|}{\texttt{emEU}} & 0.6501{\scriptsize $\pm$0.2217} & 0.5842{\scriptsize $\pm$0.1941} & $<$0.0001 \\
\multicolumn{2}{c|}{\texttt{emER}} & 0.7890{\scriptsize $\pm$0.2499} & 0.6014{\scriptsize $\pm$0.2331} & $<$0.0001 \\
\multicolumn{2}{c|}{\texttt{soME}} & 0.6872{\scriptsize $\pm$0.4048} & 0.6701{\scriptsize $\pm$0.4138} & 0.0834 \\
\multicolumn{2}{c|}{\texttt{soRE}} & 0.7268{\scriptsize $\pm$0.3713} & 0.5498{\scriptsize $\pm$0.3790} & $<$0.0001 \\
\multicolumn{2}{c|}{\texttt{moML}} & 0.9962{\scriptsize $\pm$0.0494} & 0.5077{\scriptsize $\pm$0.3285} & $<$0.0001 \bigstrut[b]\\
\hline
\multicolumn{2}{c|}{Avg. Purity} & 0.8034 & 0.6015 & - \bigstrut\\
\hline
\end{tabular}%

\end{adjustbox}
\end{table}
\endgroup

\begingroup
\setlength{\tabcolsep}{4.2pt}
\begin{table*}[t!]
    \centering
    \caption{\uline{Observation 2 (further analysis): Real-world group anchors can be well explained by scalar anchor strengths associated with each node, together with local competition.} 
    We can achieve high accuracy (92.4\% on average) if we predict the node with the highest anchor proportion (a scalar) as the anchor in each hyperedge.}
    \label{tab:scalar_acc}
\begin{adjustbox}{max width=\textwidth}
\begin{tabular}{c|cccccccccccccccc|c}
\hline
\multirow{2}[2]{*}{\textbf{Dataset}} & \multicolumn{2}{c}{\texttt{coAA}} & \multicolumn{2}{c}{\texttt{coDB}} & \multicolumn{2}{c}{\texttt{coSM}} & \multirow{2}[2]{*}{\texttt{qaBI}} & \multirow{2}[2]{*}{\texttt{qaPH}} & \multirow{2}[2]{*}{\texttt{qaMA}} & \multirow{2}[2]{*}{\texttt{qaST}} & \multirow{2}[2]{*}{\texttt{emEN}} & \multirow{2}[2]{*}{\texttt{emEU}} & \multirow{2}[2]{*}{\texttt{emER}} & \multirow{2}[2]{*}{\texttt{soME}} & \multirow{2}[2]{*}{\texttt{soRE}} & \multirow{2}[2]{*}{\texttt{moML}} & \multirow{2}[2]{*}{Avg.} \bigstrut[t]\\
      & (first) & (last) & (first) & (last) & (first) & (last) &       &       &       &       &       &       &       &       &       &       &  \bigstrut[b]\\
\hline
\textbf{Acc. (\%)}  & 93.5 & 92.9 & 97.1 & 96.1 & 98.3 & 100.0 & 98.9 & 98.6 & 100.0 & 100.0 & 80.6* & 59.4* & 81.3* & 92.8 & 88.8 & 100.0 & 92.4 \bigstrut\\
\hline
\multicolumn{18}{l}{\small *Email datasets, especially \texttt{emEU}, contain many repeated hyperedges consisting of the same nodes but with different anchors} \\
\end{tabular}%
\end{adjustbox}
\end{table*}
\endgroup

\subsection{Observation 2: Stable Yet Contextual Anchorship Across Groups (Tables~\ref{tab:label_purity} \& \ref{tab:scalar_acc})}\label{sec:intuition:global_consistency}

Group anchors are \textit{edge-dependent}, i.e., the same node might be the anchor in some hyperedges but not the anchor in others.
Even so, can we still find some patterns about each node's anchorship (i.e., whether the node is the anchor or not) across different hyperedges?
Our second observation provides insights into the question.

\begin{observation}\label{intu:global_consistency}
    In real-world group interactions, whether a node is the group anchor or not is overall stable, yet not fully homogeneous, across different hyperedges.
    That is, if a node is (not) the anchor in one hyperedge, it is likely, yet not always, (not) the anchor in another hyperedge as well.
\end{observation}

\smallsection{Intuition.}
Again, let us provide some concrete real-world intuition.
In co-authorship datasets, the last author of a paper is usually a professor, who is likely to be the last author of other papers, but can also collaborate with professors as a non-last author.
In email datasets, some employees are in charge of communicating with other companies, who frequently send emails to others, but also receive emails as responses.
In movie cast datasets, the leading actor in one movie is usually a famous one, who is likely to be the leading actor in other movies, but may also play supporting roles.

\begin{algorithm}[t]
    \caption{\ours: Stage 1}\label{algo:stage1}
    \SetKwInput{KwInput}{Input}
    \SetKwInput{KwOutput}{Output}
    \KwInput{\textbf{(1)} $V$ and $E$: topology;
    \textbf{(2)} $E'$ and $A(e), \forall e \in E'$: known group anchors;
    \textbf{(3)} $X$: topological feature matrix;
    \textbf{(4)} $n^{(1)}_{ep}$: number of optimization epochs
    }
    \KwOutput{$s^{(1)}_{v;e}, \forall v \in e \in E$: learned topology-based scores}
    \For{$i_{ep} = 1, 2, \ldots, n^{(1)}_{ep}$}{
        $s^{(1)}_{v;e} = \operatorname{MLP}(X; \theta)$ \Comment*[f]{MLP forward pass} \\ 
        $\mathcal{L}^{(1)} = -\sum_{e \in E'} \log \frac{\exp(s^{(1)}_{A(e);e})}{\sum_{u \in e} \exp(s^{(1)}_{u;e})}$  \Comment*[f]{Eq.~\eqref{eq:loss_stage_1}} \\
        Update $\theta$ w.r.t. $\frac{\partial \mathcal{L}^{(1)}}{\partial \theta}$ \Comment*[f]{Gradient descent} \\
    }       
 \Return $s^{(1)}_{v;e} = \operatorname{MLP}(X; \theta)$ \Comment*[f]{Trained scores}
\end{algorithm}

\smallsection{Evidence.}
Again, we provide more detailed evidence.
We examine the \textit{anchor purity} of nodes in real-world hypergraphs.

\begin{definition}[Anchor degree, anchor proportion and anchor purity]\label{def:seed_member_prop_purity}
    Given a hypergraph $H = (V, E)$, the \textit{anchor degree} $\delta_v$ of a node $v \in V$ is the number of hyperedges where $v$ is the group anchor, i.e., $\delta_v = |\mtsetbr{e \in E: A(e) = v}|$, the  
    \textit{anchor proportion} $p_v$ of $v$ is the proportion of hyperedges where $v$ is the anchor, i.e., $p_v = \frac{\delta_v}{d_v}$,
    and the \textit{anchor purity} $\rho_v$ of $v$ is the probability that $v$ is the anchor in both or neither of two hyperedges containing $v$ picked uniformly at random, i.e.,
    $\rho_v = \left({\binom{\delta_v}{2} + \binom{d_v - \delta_v}{2}}\right) / {\binom{d_v}{2}}$.
\end{definition}
When we observe that $v$ is (not) the anchor in one hyperedge, the anchor purity is the probability that $v$ is also (not) the anchor in another hyperedge.
Anchor purity is not considered for nodes appearing in only a single hyperedge.

As shown in \cref{tab:label_purity}, in many real-world hypergraphs, the average anchor purity is high, yet not near 100\%.
The average anchor purity of the nodes in each real-world dataset is significantly (with one-sided $t$-tests) higher than its randomized counterparts where the anchor is randomly chosen in each hyperedge.
Yet, the purity is not near 100\% in most cases.

\smallsection{Further analysis.}
The statistics suggest that there is global consistency and local variability at the same time, which inspires us to hypothesize that, 
(1) each node has its \textit{global anchor strength} across different hyperedges, yet
(2) there is \textit{local competition} in each hyperedge.
Indeed, if we set the anchor strength of each node as its anchor proportion (see \cref{def:seed_member_prop_purity}), and take the node with the highest anchor proportion in each hyperedge as the anchor, the known group anchors are well explained.
As shown in \cref{tab:scalar_acc}, the average accuracy is 92.4\% over all the datasets.
Note that the observation in \cref{tab:scalar_acc} is not intended to provide a realistic way to predict group anchors since we need ground-truth anchors to compute anchor proportions.
Instead, it (1) shows the \textit{existence} of such anchor strengths that can accurately indicate group anchors, and (2) validates the usefulness of considering local contextual interplay among nodes in the same group.
This further motivates us to set ``finding such anchor strengths that well explain known anchors'' as an objective, and to consider local contextual interplay in our method \ours, which we shall introduce below.

\section{The Proposed Method: \ours}\label{sec:method}

We shall introduce the proposed method \ours for group anchor identification, whose mechanisms are intuitively and explicitly based on our observations in \cref{sec:observations}.

\revisekddcolor
\smallsection{Overview.}
The proposed method \ours has two stages.
In Stage 1, based on \cref{intu:topological_feat} (topological features are informative about group anchors), we train a multilayer perceptron (MLP) to learn a \textit{topology-based score} for each node-hyperedge pair $(v, e)$.
In Stage 2, based on \cref{intu:global_consistency} (anchorship is overall stable across different hyperedges), we train a scalar \textit{anchor strength} for each node shared across all hyperedges.
Based on our further analysis of \cref{intu:global_consistency} (\cref{tab:scalar_acc}), we use \textit{global strength aggregation} to obtain an aggregated score for each node from different hyperedges.
Finally, in hyperedge, we predict the node with the highest aggregated score as the anchor.
Notably, \ours is semi-supervised, using information from hyperedges both with and without known anchors, e.g., the topological features.

\color{black}

\subsection{Stage 1: \revisekdd{MLP Training and Topology-based Scores of Node-Hyperedge Pairs}}\label{sec:method:stage1}

\smallsection{Overview.}
Based on \cref{intu:topological_feat} (topological features are informative about group anchors), we aim to train a model to exploit the correlations between topological features and anchorship.
The model is intended to be lightweight, using a simple architecture and only topological features as inputs
to fit known anchors.
\revisekdd{Specifically, we train a multilayer perceptron (MLP) to learn a \textit{topology-based score} for each node-hyperedge pair $(v, e)$, so that in each hyperedge $e$ with the ground-truth anchor $v^*$, the score of $(v^*, e)$ should be higher than that of $(v', e)$ for other nodes $v' \in e$.}
After training, the topology-based scores will be used as a reference to guide the training in Stage 2.
See Algorithm~\ref{algo:stage1} for an algorithmic overview.

\smallsection{Topological features.}
In principle, any topological features can be used.
Specifically, in our experiments, for a fair comparison, we use the same four topological features used in our baselines methods~\citep{choe2023classification,zheng2024co} (see \cref{sec:exp_settings}):
\textbf{(1) node degree} (see \cref{sec:prelim}),
\textbf{(2\&3) eigenvector centrality}~\citep{bonacich1987power} and \textbf{PageRank centrality}~\citep{page1999pagerank} computed on the weighted clique expansion, and \textbf{(4) coreness}~\citep{sun2020fully,seidman1983network}.
Specifically, given input topology $V$ and $E$, to each node-hyperedge pair $(v, e)$ with $v \in e$, we associate a 33-dimensional feature vector:
\begin{itemize}[leftmargin=*]
    \item \textbf{Global features (24 dimensions).} For each of the four node centrality measures, we normalize and/or aggregate it in six different ways: \textbf{(1)} min-max normalization, \textbf{(2)} rank normalization, and \textbf{(3-6)} four different aggregations: means and standard deviations of local min-max and rank normalization.
    For each node, its global features are shared across all the hyperedges that contain the node.
    \item \textbf{Local features (9 dimensions).}
    For each of the four node centrality measures, we normalize it locally in each hyperedge $e$ in two different ways: \textbf{(1)} local min-max normalization and \textbf{(2)} local rank normalization.
    We also include the size of $e$, which is a scalar, as a local feature for all the nodes in $e$.
    For each node, its local features can be distinct in different hyperedges, which allows flexibility in modeling the edge-dependent nature of group anchors.
\end{itemize}
We construct a topological feature matrix $X$ by collecting the 33-dimensional feature vectors $X_{v;e}$ for all node pairs.
See Appendix~\ref{appx:feature_process} for more details on the features. 

\smallsection{\revisekdd{MLP training and topology-based scores.}}
We use an MLP parameterized with $\theta$ to obtain the \textit{topology-based scores} $s^{(1)}_{v;e}$ of each node-hyperedge pair $(v, e)$: 
    $s^{(1)}_{v;e} = \operatorname{MLP}(X; \theta)$,
with loss function
\begin{equation}\label{eq:loss_stage_1}
    \mathcal{L}^{(1)} = -\sum\nolimits_{e \in E'} \log 
    \left(\exp\left(s^{(1)}_{A(e);e}\right) \middle/ \sum\nolimits_{u \in e} \exp\left(s^{(1)}_{u;e}\right)\right),
\end{equation}
where $E'$ is the set of hyperedges with known anchors and $A(e)$ is the anchor in each $e \in E'$ (see \cref{prob:seed_identification}).
Minimizing the loss function encourages a large (relative) score $s^{(1)}_{A(e);e}$ of the anchor compared to $s^{(1)}_{u;e}$ for other nodes $u \in e$.

\begin{algorithm}[t]
    \caption{\ours: Stage 2}\label{algo:stage2}
    \SetKwInput{KwInput}{Input}
    \SetKwInput{KwOutput}{Output}
    \KwInput{\textbf{(1)} $V$ and $E$: topology;
    \textbf{(2)} $E'$ and $A(e), \forall e \in E'$: known group anchors;
    \textbf{(3)} $s^{(1)}_{v;e}, \forall v \in e \in E$: learned strengths from Stage 1;
    \textbf{(4)} $\alpha^{(2)}$: loss term coefficient;
    \textbf{(5)} $w^{(2)}$: global aggregation weight;
    \textbf{(6)} $n^{(2)}_{ep}$: number of optimization epochs}
    \KwOutput{$\tilde{A}(e), \forall e \in E \setminus E'$: predicted group anchors}
    $s^{(2)}_{v} \gets 1, \forall v \in V$ \Comment*[f]{Initialization} \\
    \For{$i_{ep} = 1, 2, \ldots, n^{(2)}_{ep}$}{
        $\mathcal{L}^{(2)}_1 = -\sum_{e \in E'} \log \frac{\exp(s^{(2)}_{A(e)})}{\sum_{u \in e} \exp(s^{(2)}_u)}$ \Comment*[f]{Eq.~\eqref{eq:loss_stage_2_label}} \\
        $\mathcal{L}^{(2)}_2 = -\sum_{e \in E'}
    \frac{\exp(s^{(2)}_{A(e)})}{\sum_{u \in e} \exp(s^{(2)}_u)} \cdot \frac{\exp(s^{(1)}_{A(e);e})}{\sum_{u \in e} \exp(s^{(1)}_{u;e})}$ \Comment*[f]{Eq.~\eqref{eq:loss_stage_2_align}} \\
        Update each $s^{(2)}_{v}$ w.r.t. $\frac{\partial \left(\mathcal{L}^{(2)}_1 + \alpha^{(2)} \mathcal{L}^{(2)}_2\right)}{\partial s^{(2)}_{v}}$ \Comment*[f]{Gradient descent} \\
    }
    $\hat{A}(e) = \arg \max_{v^* \in e} s^{(2)}_{v^*}, \forall e \in E$ \Comment*[f]{Max anchor strength} \\         
    $\hat{p}_v = \frac{w^{(2)} 
    \sum_{e \in E'} \mathbf{1}[A(e) = v]
    + \sum_{e \in E \setminus E'} \mathbf{1}[\hat{A}(e) = v]}{d_v}, \forall v \in V$ \Comment*[f]{Eq.~\eqref{eq:pred_member_prop}} \\    
 \Return $\tilde{A}(e) = \arg\max_{v^* \in e} \hat{p}_{v^*}, \forall e \in E \setminus E'$ \Comment*[f]{Final prediction}
\end{algorithm}

\subsection{\revisekdd{Stage 2: Anchor Strength Learning, Global Aggregation, and Final Prediction}}\label{sec:method:stage2}

\smallsection{Overview.}
Based on \cref{intu:global_consistency}, we aim to find anchor strengths (scalars indicating the overall likelihood that each node is the anchor), together with local competition and global aggregation, to explain the known anchors.
\revisekdd{Specifically, for each node, we train a scalar \textit{anchor strength} shared across all hyperedges, 
so that 
(1) in each hyperedge $e$ with the ground-truth anchor $v^*$, the anchor strength of $v^*$ should be higher than that of other nodes $v' \in e$, and
(2) based on \cref{intu:topological_feat} again, the relative relations of the strengths in each hyperedge should align well with the topology-based scores in Stage 1.
Then, based on our further analysis of \cref{intu:global_consistency} (\cref{tab:scalar_acc}), we use \textit{global strength aggregation}, i.e., compute the anchor proportion (see \cref{def:seed_member_prop_purity}) assuming the predictions based on the learned anchor strengths are all correct, to further enhance the prediction robustness and accuracy.
Finally, in hyperedge, we predict the node with the highest aggregated score as the anchor.}
See Algorithm~\ref{algo:stage2} for an algorithmic overview.

\smallsection{\revisekdd{Anchor strength learning.}}
Let $s^{(2)}_v$ be the learnable anchor strength of each node $v$.
The loss function consists of two parts.
The first part, similar to that in Stage 1, aims to fit the known anchors:
\begin{equation}\label{eq:loss_stage_2_label}
    \mathcal{L}^{(2)}_1 = -\sum\nolimits_{e \in E'} \log \left({\exp\left(s^{(2)}_{A(e)}\right)} \middle/ {\sum\nolimits_{u \in e} \exp\left(s^{(2)}_u\right)}\right).
\end{equation}
The second part aligns with the learned scores from Stage 1:
\begin{equation}\label{eq:loss_stage_2_align}
    \mathcal{L}^{(2)}_2 = -\sum\nolimits_{e \in E'}
    \frac{\exp\left(s^{(2)}_{A(e)}\right)}{\sum\nolimits_{u \in e} \exp\left(s^{(2)}_u\right)} \cdot \frac{\exp\left(s^{(1)}_{A(e);e}\right)}{\sum\nolimits_{u \in e} \exp\left(s^{(1)}_{u;e}\right)},
\end{equation}
which measures the distance between $s^{(1)}_{v;e}$'s and $s^{(2)}_v$'s after within-edge normalization.
The final loss function is $\mathcal{L}^{(2)} = \mathcal{L}^{(2)}_1 + \alpha^{(2)} \mathcal{L}^{(2)}_2$ with a hyperparameter coefficient $\alpha^{(2)}$.

\smallsection{Global strength aggregation.}
Recall the observation of the high accuracy of anchor proportions (see \cref{tab:scalar_acc} in \cref{sec:intuition:global_consistency}).
The anchor proportion of a node $v$ can be seen as the global aggregation of its anchor indicators $\mathbf{1}[A(e) = v]$.\footnote{$\mathbf{1}[A(e) = v] = 1$ if $A(e) = v$, and $\mathbf{1}[A(e) = v] = 0$ otherwise.}
Inspired by this, after learning anchor strengths $s^{(2)}_v$'s, we first find the node $\hat{A}(e)$ with the highest learned strengths in each hyperedge $e$, i.e., $\hat{A}(e) = \arg \max_{v^* \in e} s^{(2)}_{v^*}, \forall e \in E$ (we randomly pick one if multiple nodes have the highest learned anchor strength), and compute the predicted anchor proportion $\hat{p}_v$ of each node $v$:
\begin{equation}\label{eq:pred_member_prop}
\small
    \hat{p}_v = \frac{1}{d_v} \left({w^{(2)} 
    \sum\nolimits_{e \in E'} \mathbf{1}[A(e) = v]
    + \sum\nolimits_{e \in E \setminus E'} \mathbf{1}[\hat{A}(e) = v]}\right),
\end{equation}
where we use a hyperparameter $w^{(2)}$ as the global aggregation weight on training (i.e., ground-truth) labels.
Such global aggregation considers the predictions across different hyperedges and increases the prediction robustness~\citep{kumar2018rev2}.

\smallsection{\revisekdd{Final prediction.}}
Finally, in each hyperedge, we predict the node with the highest aggregated score $\hat{p}_v$ as the anchor.

\begingroup
\setlength{\tabcolsep}{2.5pt}
\begin{table*}[t!]
    \centering
    \caption{\uline{Q1: \ours achieves higher accuracy in identifying group anchors than all the baselines in most cases.} For each setting, the best performance is highlighted in bold, while the second-best is underlined.
    The mean accuracy values (\%) over five random splits are reported with standard deviations. OOM represents ``out of memory''.
    \revisekdd{\ours also outperforms the baseline methods with different training ratios (see Appendix~\ref{appx:diff_training_ratios}) and w.r.t. other evaluation metrics (see Appendix~\ref{appx:diff_eval_metrics}).}}
    \label{tab:single_node_res}
\begin{adjustbox}{max width=\textwidth}
\begin{tabular}{c|cccccccccccccccc}
\hline
\multirow{2}[2]{*}{\textbf{Dataset}} 
& \multicolumn{2}{c}{\texttt{coAA}} 
& \multicolumn{2}{c}{\texttt{coDB}} 
& \multicolumn{2}{c}{\texttt{coSM}} 
& \multirow{2}[2]{*}{\texttt{qaBI}} 
& \multirow{2}[2]{*}{\texttt{qaPH}} 
& \multirow{2}[2]{*}{\texttt{qaMA}} 
& \multirow{2}[2]{*}{\texttt{qaST}} 
& \multirow{2}[2]{*}{\texttt{emEN}} 
& \multirow{2}[2]{*}{\texttt{emEU}} 
& \multirow{2}[2]{*}{\texttt{emER}} 
& \multirow{2}[2]{*}{\texttt{soME}} 
& \multirow{2}[2]{*}{\texttt{soRE}} 
& \multirow{2}[2]{*}{\texttt{moML}} \bigstrut[t]\\
      
& (first) 
      & (last) 
      & (first) 
      & (last) 
      & (first) 
      & (last) 
      &       
      &       
      &       
      &       
      &       
      &       
      &       
      &       
      &       
      &  \bigstrut[b]\\
    \hline
    WHATsNet & \uline{45.2{\scriptsize$\pm$0.2}} & \uline{45.8{\scriptsize$\pm$0.3}} & \uline{42.5{\scriptsize$\pm$0.3}} & 45.4{\scriptsize$\pm$0.2} & 34.3{\scriptsize$\pm$2.0} & \uline{39.8{\scriptsize$\pm$1.8}} & \uline{85.6{\scriptsize$\pm$0.4}} & \uline{88.1{\scriptsize$\pm$0.1}} & \uline{35.8{\scriptsize$\pm$1.1}} & \uline{31.2{\scriptsize$\pm$0.2}} & \uline{50.8{\scriptsize$\pm$3.4}} & 51.0{\scriptsize$\pm$0.3} & \uline{66.6{\scriptsize$\pm$3.1}} & \textbf{75.5{\scriptsize$\pm$0.2}} & 97.4{\scriptsize$\pm$0.3} & 41.4{\scriptsize$\pm$0.3} \bigstrut[t]\\
    CoNHD-U & 42.7{\scriptsize$\pm$1.3} & 42.2{\scriptsize$\pm$2.0} & 41.2{\scriptsize$\pm$0.1} & 42.7{\scriptsize$\pm$0.3} & 31.4{\scriptsize$\pm$1.9} & 37.9{\scriptsize$\pm$2.1} & 78.7{\scriptsize$\pm$0.5} & 76.0{\scriptsize$\pm$1.0} & 29.0{\scriptsize$\pm$4.2} & 25.4{\scriptsize$\pm$3.8} & 44.0{\scriptsize$\pm$4.3} & \textbf{52.8{\scriptsize$\pm$0.3}} & 65.3{\scriptsize$\pm$2.2} & 74.3{\scriptsize$\pm$0.3} & 96.8{\scriptsize$\pm$0.6} & 42.4{\scriptsize$\pm$0.5} \\
    CoNHD-I & 44.5{\scriptsize$\pm$0.8} & 44.7{\scriptsize$\pm$0.3} & 41.4{\scriptsize$\pm$0.5} & 43.5{\scriptsize$\pm$0.7} & 29.8{\scriptsize$\pm$2.6} & 39.4{\scriptsize$\pm$1.9} & 79.3{\scriptsize$\pm$0.4} & 77.3{\scriptsize$\pm$0.2} & 29.8{\scriptsize$\pm$6.1} & 26.6{\scriptsize$\pm$1.3} & 45.1{\scriptsize$\pm$3.8} & \uline{52.4{\scriptsize$\pm$0.4}} & 64.6{\scriptsize$\pm$1.2} & 74.6{\scriptsize$\pm$0.4} & \uline{97.5{\scriptsize$\pm$0.6}} & \uline{42.7{\scriptsize$\pm$0.3}} \\
    HNHN  & 39.7{\scriptsize$\pm$0.0} & 41.2{\scriptsize$\pm$0.0} & 35.5{\scriptsize$\pm$0.4} & 39.1{\scriptsize$\pm$0.4} & 33.2{\scriptsize$\pm$1.4} & 33.7{\scriptsize$\pm$0.6} & 63.5{\scriptsize$\pm$1.4} & 37.7{\scriptsize$\pm$0.1} & 30.5{\scriptsize$\pm$0.8} & 22.9{\scriptsize$\pm$0.1} & 35.8{\scriptsize$\pm$2.1} & 49.2{\scriptsize$\pm$1.2} & 41.8{\scriptsize$\pm$6.4} & 56.4{\scriptsize$\pm$0.4} & 53.4{\scriptsize$\pm$0.8} & 35.2{\scriptsize$\pm$0.3} \\
    HGNN  & 44.1{\scriptsize$\pm$0.0} & 45.9{\scriptsize$\pm$0.1} & 41.9{\scriptsize$\pm$0.1} & 44.6{\scriptsize$\pm$0.3} & 33.1{\scriptsize$\pm$0.3} & 38.1{\scriptsize$\pm$0.6} & 81.7{\scriptsize$\pm$0.3} & 74.9{\scriptsize$\pm$0.8} & 28.9{\scriptsize$\pm$1.0} & 30.4{\scriptsize$\pm$0.5} & 40.1{\scriptsize$\pm$0.7} & 49.3{\scriptsize$\pm$0.2} & 42.0{\scriptsize$\pm$0.9} & 62.1{\scriptsize$\pm$2.1} & 84.6{\scriptsize$\pm$3.4} & 37.9{\scriptsize$\pm$0.5} \\
    HCHA  & 38.9{\scriptsize$\pm$0.2} & 39.4{\scriptsize$\pm$0.4} & 35.3{\scriptsize$\pm$0.6} & 31.4{\scriptsize$\pm$0.7} & 33.2{\scriptsize$\pm$1.1} & 35.2{\scriptsize$\pm$3.6} & 69.8{\scriptsize$\pm$2.1} & 68.0{\scriptsize$\pm$1.5} & 31.0{\scriptsize$\pm$2.4} & 23.4{\scriptsize$\pm$2.6} & 18.8{\scriptsize$\pm$2.2} & 45.4{\scriptsize$\pm$0.5} & 46.0{\scriptsize$\pm$4.9} & 30.7{\scriptsize$\pm$2.2} & 52.7{\scriptsize$\pm$0.6} & 17.3{\scriptsize$\pm$0.4} \\
    HAT   & 43.5{\scriptsize$\pm$0.3} & 45.8{\scriptsize$\pm$0.1} & 38.1{\scriptsize$\pm$1.4} & 40.5{\scriptsize$\pm$2.0} & 30.1{\scriptsize$\pm$0.5} & 33.0{\scriptsize$\pm$1.0} & 75.8{\scriptsize$\pm$0.3} & 81.3{\scriptsize$\pm$0.2} & 29.2{\scriptsize$\pm$1.2} & 23.9{\scriptsize$\pm$0.2} & 49.8{\scriptsize$\pm$1.7} & 50.8{\scriptsize$\pm$0.4} & 42.3{\scriptsize$\pm$7.0} & 68.0{\scriptsize$\pm$0.9} & 92.4{\scriptsize$\pm$0.9} & 36.1{\scriptsize$\pm$0.8} \\
    UniGCN & 43.3{\scriptsize$\pm$0.5} & 45.8{\scriptsize$\pm$0.4} & 41.2{\scriptsize$\pm$0.7} & \uline{45.8{\scriptsize$\pm$0.6}} & \uline{34.8{\scriptsize$\pm$2.7}} & 39.2{\scriptsize$\pm$4.2} & 76.3{\scriptsize$\pm$1.2} & 78.0{\scriptsize$\pm$1.4} & 35.0{\scriptsize$\pm$3.4} & 30.7{\scriptsize$\pm$0.4} & 45.3{\scriptsize$\pm$2.7} & 49.6{\scriptsize$\pm$0.7} & 55.5{\scriptsize$\pm$2.7} & 68.9{\scriptsize$\pm$1.3} & 88.1{\scriptsize$\pm$0.6} & 40.1{\scriptsize$\pm$1.8} \\
    HNN   & 37.7{\scriptsize$\pm$0.1} & 39.3{\scriptsize$\pm$0.1} & 31.4{\scriptsize$\pm$0.8} & 36.3{\scriptsize$\pm$1.2} & 32.7{\scriptsize$\pm$0.7} & 36.8{\scriptsize$\pm$0.8} & 63.8{\scriptsize$\pm$0.9} & 62.6{\scriptsize$\pm$0.6} & 29.2{\scriptsize$\pm$3.2} & 24.8{\scriptsize$\pm$0.5} & 38.7{\scriptsize$\pm$2.7} & OOM   & 47.1{\scriptsize$\pm$4.3} & 56.0{\scriptsize$\pm$0.6} & 57.2{\scriptsize$\pm$7.6} & 33.8{\scriptsize$\pm$0.6} \bigstrut[b]\\
    \hline
    \ours & \textbf{49.7{\scriptsize$\pm$0.1}} & \textbf{50.6{\scriptsize$\pm$0.0}} & \textbf{46.5{\scriptsize$\pm$0.2}} & \textbf{49.9{\scriptsize$\pm$0.1}} & \textbf{40.9{\scriptsize$\pm$0.7}} & \textbf{48.1{\scriptsize$\pm$1.3}} & \textbf{87.4{\scriptsize$\pm$0.2}} & \textbf{88.7{\scriptsize$\pm$0.1}} & \textbf{40.6{\scriptsize$\pm$5.3}} & \textbf{36.6{\scriptsize$\pm$0.4}} & \textbf{53.6{\scriptsize$\pm$2.4}} & 50.9{\scriptsize$\pm$0.2} & \textbf{67.8{\scriptsize$\pm$2.6}} & \uline{74.9{\scriptsize$\pm$0.6}} & \textbf{97.8{\scriptsize$\pm$0.3}} & \textbf{45.0{\scriptsize$\pm$0.3}} \bigstrut\\
    \hline
\end{tabular}%
\end{adjustbox}
\end{table*}
\endgroup

\subsection{Complexity Analysis}\label{sec:method:complexity}

Due to its simplicity, \ours has low complexity, especially time complexity and the number of learnable parameters.
For the whole process of \ours,
the training time complexity is
$O(\sum_{e \in E} (n^{(1)}_{ep} |e| n_{f} D_h + n^{(2)}_{ep}))$,
the inference time complexity is
$O(\sum_{e \in E} |e| n_{f} D_h)$,
the space complexity is
$O(n_{f} (D_h + \sum_{e \in E} |e|))$, and
the number of learnable parameters is
$O(D_h (n_{f} + 2) + 1 + |V|)$.

\smallsection{Notes.}
In our experiments, we use $D_h = 64$, $n_f = 33$ (see \cref{sec:method:stage1}), and $n^{(1)}_{ep} = n^{(2)}_{ep} = 100$.
For $|V|$ and $|e|$, see \cref{tab:dataset_summary}.
See Appendix~\ref{appx:complexity} for the detailed analysis.

\section{Experiments}\label{sec:experiments}

In this section, we present the experimental results to validate the empirical superiority of the proposed method \ours.
We evaluate \ours by answering the following questions:
\begin{itemize}[leftmargin=*]
    \item \textbf{Q1.} Does \ours accurately predict group anchors?
    \item \textbf{Q2.} How efficient is \ours compared to baselines?
    \item \textbf{Q3.} Does each algorithmic design of \ours make a meaningful contribution to the performance?
    \item \textbf{Q4.} Is \ours useful in downstream applications?
\end{itemize}

\subsection{Experimental Settings}\label{sec:exp_settings}

\smallsection{Data processing.}
For each dataset, we split the hyperedges into three categories (training, validation, and test) based on unique hyperedges in $E^*$, to ensure that the same group of nodes does not appear in two different categories.
We focus on the scenarios with limited training data (i.e., known group anchors), which is common in the real world with label scarcity~\citep{zhu2022introduction,zhu2005semi,yang2022semi}.
The ratios of unique hyperedges for training, validation, and test are 7.5\%, 2.5\%, and 90\%, respectively.
We obtain five random splits for each dataset, and the reported results are averaged over the five splits. 
See \cref{sec:observations} for the basic information of the thirteen real-world datasets.

\smallsection{\ours.}
We use a two-layer MLP with hidden dimension 64 in Stage 1.
We fine-tune the learning rates in both stages and the loss term coefficient and global aggregation weight in Stage 2 (see \cref{sec:method:stage2}).
We use Adam~\citep{kingma2014adam} as the optimizer in both stages.

\smallsection{Baseline methods.}
To the best of our knowledge, we are the first to consider the problem of group anchor identification, so no immediate baselines exist.
As mentioned in \cref{sec:rel_wk}, group anchor identification can be seen as a special case of edge-dependent node classification (ENC), and we adapt existing methods for ENC with proper modification as baseline methods:
\textbf{(1)} \textbf{WHATsNet}~\citep{choe2023classification},
\textbf{(2\&3)} \textbf{CoNHD-U} and \textbf{CoNHD-I}, two variants of CoNHD~\citep{zheng2024co},
and \textbf{(4-9)} 
\textbf{HNHN}~\citep{dong2020hnhn}, 
\textbf{HGNN}~\citep{feng2019hypergraph}, 
\textbf{HCHA}~\citep{bai2021hypergraph}, 
\textbf{HAT}~\citep{hwang2021hyfer}, 
\textbf{UniGCN}~\citep{huang2021unignn}, and 
\textbf{HNN}~\citep{aponte2022hypergraph}, 
six baselines used by~\citet{choe2023classification}.
We train the baseline methods on the original ENC problem, and follow the original papers~\citep{choe2023classification,zheng2024co} for their settings (e.g., hyperparameter tuning).
For all baseline methods, additional label information is provided (e.g., the first and last authors of each paper are both provided).
Each baseline predicts a label distribution for each node-hyperedge pair, and we pick the node with the highest score of the label corresponding to group anchors in each hyperedge.
See Appendix~\ref{appx:exp_settings} for more details on experimental settings.

\subsection{Q1. Accuracy (Tables~\ref{tab:single_node_res} and \ref{tab:res_multi_main})}\label{sec:experiments:acc}
We evaluate the accuracy of group anchor identification of \ours and the baselines on thirteen real-world datasets.
The accuracy is the proportion of hyperedges where the predicted group anchor is correct.
For each method and each setting, we report the average accuracy over five random splits (see \cref{sec:exp_settings}) with the standard deviation.
As shown in \cref{tab:single_node_res}, \ours achieves higher accuracy than all the baselines in most cases.

Overall, WHATsNet~\citep{choe2023classification} is the strongest baseline method, performing comparably with \ours on several datasets, especially \texttt{emEU}, \texttt{soME}, and \texttt{soRE}.
In our understanding, compared to \ours, the baseline methods are heavily parameterized (to be discussed in detail below in \cref{sec:experiments:efficiency}; see also the complexities of \ours in \cref{sec:method:complexity}), which makes the training of such methods difficult and prone to overfitting~\citep{ying2019overview} in our experiments with (1) label scarcity~\citep{karystinos2000overfitting}, i.e., only 7.5\% hyperedges contain known anchors, and (2) imbalanced labels~\citep{li2020analyzing}, i.e., a single anchor is in each hyperedge.
\revisekdd{The performance of \ours is relatively weak on \texttt{emEU}, possibly because our observations on \texttt{emEU} are also weaker (see Tables~\ref{tab:degree_perf} to \ref{tab:scalar_acc}).}

\smallsection{Additional features.}
As discussed in Sections~\ref{sec:prelim} and \ref{sec:problem_state}, attributes can be incorporated into our method \ours if they are given.
We conduct additional experiments using additional node features from the metadata of the \texttt{coAA} dataset for each node (author): the number of publications, the number of citations, the h-index, and the p-index.
With such additional features, the accuracy of \ours is further improved, increasing from 49.7\% to 50.8\% for first authors, and from 50.6\% to 51.8\% for last authors.

\begingroup
\setlength{\tabcolsep}{7pt}
\begin{table}[t!]
    \centering
    \caption{
    \uline{In scenarios with multiple anchors in each group, \ours consistently achieves higher accuracy in identifying group anchors than all the baseline.}
    For each setting, the best performance is highlighted in bold, while the second-best is underlined.
    The mean accuracy values (\%) over five random splits are reported with standard deviations.}
    \label{tab:res_multi_main}
\begin{adjustbox}{max width=\textwidth}
    \begin{tabular}{c|ccc}
    \hline
    \textbf{Dataset} & \texttt{coAA} & \texttt{coDB} & \texttt{coSM} \bigstrut\\
    \hline
    WHATsNet & 31.87 & 31.71 & 20.56 \bigstrut[t]\\
    CoNHD-U & \uline{34.03} & 32.97 & 19.57 \\
    CoNHD-I & 33.86 & \uline{33.06} & \uline{21.62} \\
    HNHN  & 26.88 & 25.72 & 21.01 \\
    HGNN  & 32.25 & 30.69 & 21.13 \\
    HCHA  & 25.58 & 24.10 & 23.30 \\
    HAT   & 29.57 & 24.25 & 20.02 \\
    UniGCN & 25.78 & 31.12 & 19.74 \\
    HNN   & 22.95 & 22.10 & 21.11 \bigstrut[b]\\
    \hline
    \ours & \textbf{37.50} & \textbf{37.42} & \textbf{32.81} \bigstrut\\
    \hline
    \end{tabular}%
\end{adjustbox}
\end{table}
\endgroup

\smallsection{Multiple anchors.}
As discussed in Sections~\ref{sec:intro} and \ref{sec:problem_state}, the scenario with a single anchor in each group is common.
To the best of our knowledge, no real-world datasets with multiple anchors are available, so we consider
co-authorship datasets and set both the first and second authors as anchors (motivated by ``equal contribution'').
As shown in Table~\ref{tab:res_multi_main}, \ours still outperforms the baseline methods.
\begin{figure}[t!]
    \centering
    \includegraphics[width=\linewidth]{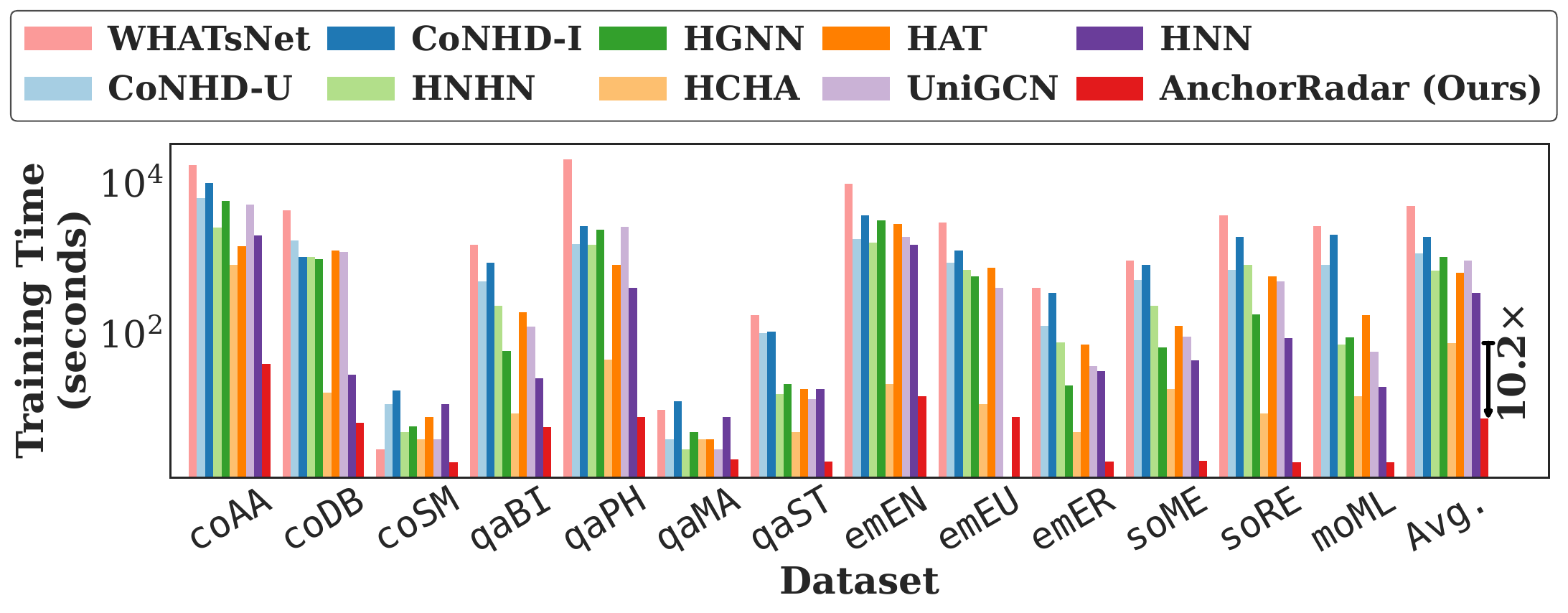} \\
    \includegraphics[width=\linewidth]{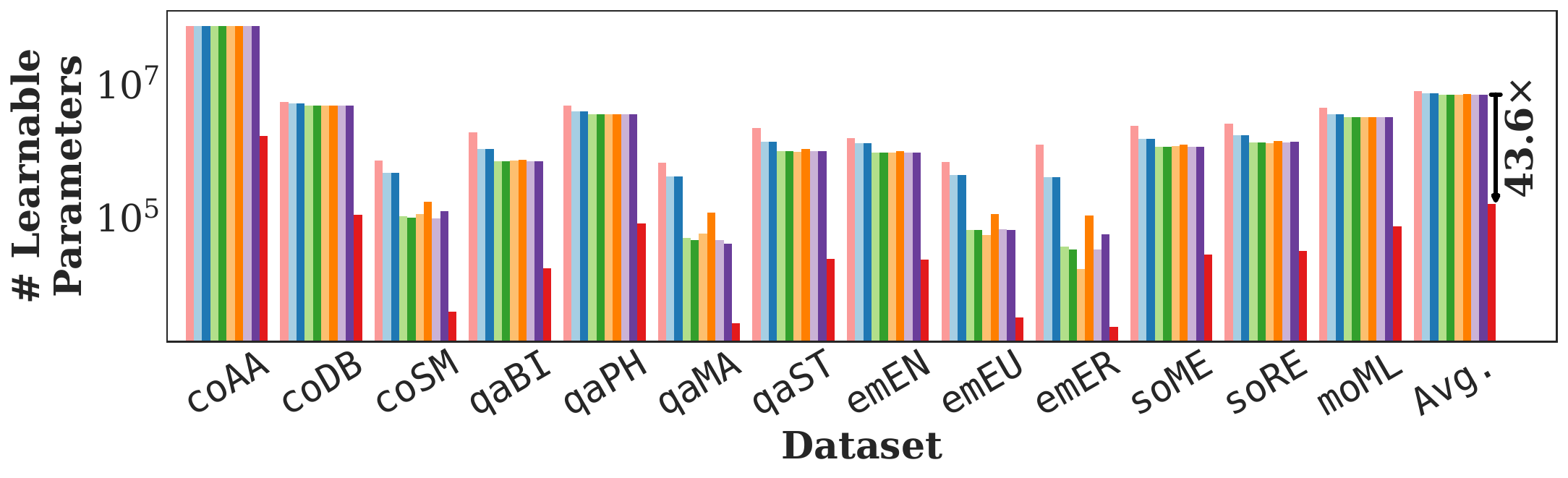}    
    \caption{\uline{Q2: \ours demands significantly less training time and uses significantly fewer learnable parameters.}
    On average, \ours uses 10.2$\times$ less training time than the fastest baseline and uses
    43.6$\times$ fewer learnable parameters than the most lightweight baseline.}
    \vspace{-1mm}
    \label{fig:training_time}    
\end{figure}

\revisekddcolor
\smallsection{Additional results.}
\ours also outperforms the baseline methods with different training ratios (see Appendix~\ref{appx:diff_training_ratios}) and w.r.t. other evaluation metrics (see Appendix~\ref{appx:diff_eval_metrics}).
The performance of \ours is robust to random seeds (see Appendix~\ref{appx:random_seeds}) and hyperparameters (see Appendix~\ref{appx:hyperparam_sens}).
When trained and tested on different datasets in the same domain, it is possible for \ours to transfer knowledge on the correlations between topological features and group anchors learned by the MLP in Stage 1 (see Appendix~\ref{appx:experiments:acc}).
\color{black}

\subsection{Q2. Training Time and Parameters (Figure~\ref{fig:training_time})}\label{sec:experiments:efficiency}
We measure the training time and the number of learnable parameters on each dataset, for \ours and the baselines.
As shown in \cref{fig:training_time}, \ours is more efficient than the baselines, using significantly less training
time and significantly fewer learnable parameters.
On average, \ours uses 10.2$\times$ less training time than the fastest baseline and uses 43.6$\times$ fewer learnable parameters than the most lightweight baseline, empirically validating the low complexities of \ours (see \cref{sec:method:complexity}).

\begingroup
\setlength{\tabcolsep}{5
pt}
\begin{table*}[t!]
    \centering
    \caption{\uline{Q3: Each component in \ours positively contributes the performance.}  The original \ours overall performs better than all the variants with some components absent. For each setting, the best performance is highlighted in bold, while the second-best is underlined.
    The mean accuracy values (\%) over five random splits are reported.}
    \label{tab:ablation_study}
\begin{adjustbox}{max width=\textwidth}
\begin{tabular}{c|cccccccccccccccc|c}
\hline
\multirow{2}[2]{*}{\textbf{Dataset}} & \multicolumn{2}{c}{\texttt{coAA}} & \multicolumn{2}{c}{\texttt{coDB}} & \multicolumn{2}{c}{\texttt{coSM}} & \multirow{2}[2]{*}{\texttt{qaBI}} & \multirow{2}[2]{*}{\texttt{qaPH}} & \multirow{2}[2]{*}{\texttt{qaMA}} & \multirow{2}[2]{*}{\texttt{qaST}} & \multirow{2}[2]{*}{\texttt{emEN}} & \multirow{2}[2]{*}{\texttt{emEU}} & \multirow{2}[2]{*}{\texttt{emER}} & \multirow{2}[2]{*}{\texttt{soME}} & \multirow{2}[2]{*}{\texttt{soRE}} & \multirow{2}[2]{*}{\texttt{moML}} & \multirow{2}[2]{*}{Avg.} \bigstrut[t]\\
        & (first) & (last) & (first) & (last) & (first) & (last) &       &       &       &       &       &       &       &       &       &       &  \bigstrut[b]\\
\hline
Stage 1 & 47.53 & 47.65 & 43.42 & 45.84 & 35.75 & 43.38 & 85.52 & 87.15 & 39.24 & 31.20 & 47.14 & 49.08 & 66.01 & 72.83 & 88.94 & 41.88 & 54.54 \bigstrut[t]\\
Stage 2 & 45.83 & 44.27 & 42.67 & 42.71 & 37.14 & 39.83 & 83.85 & 85.41 & 39.32 & 29.18 & 52.21 & 50.73 & 63.32 & 74.27 & \textbf{97.82} & 43.41 & 54.50 \\
No GA & 49.50 & \textbf{50.68} & 45.83 & \uline{49.90} & 40.12 & \uline{48.13} & 86.47 & 87.78 & \textbf{41.03} & 35.96 & 53.07 & \textbf{51.06} & \uline{67.63} & 73.67 & 96.03 & \textbf{44.99} & 57.62 \\
No LF & \uline{49.64} & 50.54 & \textbf{46.63} & 49.86 & \uline{40.53} & \textbf{48.32} & \uline{87.26} & \uline{88.69} & 40.08 & \uline{36.48} & \textbf{54.24} & 50.87 & 66.96 & \textbf{74.91} & \textbf{97.82} & 43.32 & \uline{57.89} \bigstrut[b]\\
\hline
\ours & \textbf{49.68} & \uline{50.60} & \uline{46.55} & \textbf{49.95} & \textbf{40.92} & 48.08 & \textbf{87.41} & \textbf{88.74} & \uline{40.59} & \textbf{36.57} & \uline{53.55} & \uline{50.89} & \textbf{67.77} & \uline{74.87} & \textbf{97.82} & \textbf{44.99} & \textbf{58.06} \bigstrut\\
\hline
\end{tabular}%
\end{adjustbox}
\end{table*}
\endgroup

\begingroup
\setlength{\tabcolsep}{6pt}
\begin{table*}[t!]
    \centering
    \caption{\uline{Q4: The anchor strengths from \ours provide useful information in group interaction prediction.} With additional information derived from anchor strengths, VilLain achieves higher accuracy (\%) in distinguishing realistic and fake group interactions. For each setting, the better performance is highlighted in bold.}
    \label{tab:downstream}
\begin{adjustbox}{max width=\textwidth}
\begin{tabular}{l|ccccccccccccc|c}
\hline
\multirow{2}[1]{*}{\textbf{Dataset}} & \multicolumn{2}{c}{\texttt{coDB}} & \multicolumn{2}{c}{\texttt{coSM}} & \multirow{2}[1]{*}{\texttt{qaBI}} & \multirow{2}[1]{*}{\texttt{qaPH}} & \multirow{2}[1]{*}{\texttt{qaST}} &  \multirow{2}[1]{*}{\texttt{emEN}} & \multirow{2}[1]{*}{\texttt{emEU}} & \multirow{2}[1]{*}{\texttt{emER}} & \multirow{2}[1]{*}{\texttt{soME}} & \multirow{2}[1]{*}{\texttt{soRE}} & \multirow{2}[1]{*}{\texttt{moML}} & \multirow{2}[1]{*}{\texttt{Avg.}}  \bigstrut[t]\\
 & (first) & (last) & (first) & (last) & & & & & & & & & & \bigstrut[b]\\
\hline
Original VilLain & 89.71 & 89.71 & 91.40 & 91.40 & 71.68 & 74.58 & \textbf{76.69} & 89.65 & 87.22 & 87.45 & 96.76 & 94.39 & 95.56 & 87.40 \bigstrut[t]\\
+ Anchor Strengths & \textbf{93.11} & \textbf{93.41} & \textbf{91.94} & \textbf{92.47} & \textbf{80.45} & \textbf{85.64} & 74.06 & \textbf{97.57} & \textbf{92.01} & \textbf{90.73} & \textbf{97.87} & \textbf{95.82} & \textbf{96.52} & \textbf{90.89}  \bigstrut[b]\\
\hline
\multicolumn{15}{l}{\small *The largest dataset (\texttt{coAA}) is omitted since VilLain runs out of memory on it} \\
\multicolumn{15}{l}{\small *The smallest dataset (\texttt{qaMA}) is omitted since it is too sparse and we cannot obtain enough meaningful test hyperedges} \\
\end{tabular}%
\end{adjustbox}
\end{table*}
\endgroup

\subsection{Q3. Ablation Studies (Table~\ref{tab:ablation_study})}
We evaluate the performance of different variants of \ours with different algorithmic components absent.
Specifically, we consider the following four variants:
\textbf{(1) Stage 1 only}, predicting the node with the highest learned strengths $s^{(1)}_{v;e}$ in each hyperedge,
\textbf{(2) Stage 2 only}, without fitting to the strengths learned in Stage 1 (i.e., the loss term coefficient $\alpha^{(2)} = 0$),
\textbf{(3) without global aggregation} (GA; see \cref{sec:method:stage2}) in Stage 2, and
\textbf{(4) without local features} (LF; see \cref{sec:method:stage1}) in Stage 1.
As shown in \cref{tab:ablation_study}, the performance of the original \ours is overall higher than the variants, validating that each algorithmic component in \ours positively contributes to the performance.
Specifically, the synergy between the two stages is clearly demonstrated.

\subsection{Q4. Downstream Application (Table~\ref{tab:downstream})}\label{sec:experiments:downstream}
We consider the downstream application of group-interaction prediction (see \cref{sec:intro}).
We use VilLain~\citep{lee2024villain}, a method that can obtain hyperedge embeddings without input node/hyperedge features.
For each hyperedge, we follow the original settings in~\citep{lee2024villain} to generate a fake hyperedge of the same size by randomly sampling nodes.
We use \ours to obtain learned anchor strengths $s^{(2)}_v$'s and compare the performance of VilLain with its original embeddings and with additional information from anchor strengths.
In each dataset, we take 10$\%$ hyperedges from the original test hyperedges out for the test of group-interaction prediction.
VilLain originally learns 128-dimensional hyperedge embeddings, and we generate 11 additional dimensions using anchor strengths.
As shown in \cref{tab:downstream}, with the additional dimensions derived from anchor strengths learned by \ours, VilLain achieves better accuracy in distinguishing realistic and fake group interactions.
We conjecture that there are specific patterns or distributions of anchor strengths in real-world group interactions that are distinct from random ones, which are well captured by \ours.
See Appendix~\ref{appx:downstream} for more details on the downstream-application experiments.

\section{Conclusion and     Discussions}\label{sec:conclusion}

In this work, we discuss the existence of group anchors in real-world group interactions and study the problem of identifying them (\cref{sec:problem_state}).
We discuss several observations on anchors in real-world group interactions (\cref{sec:observations}).
We propose a novel method \ours for group anchor identification (\cref{sec:method}), with intuitive algorithmic designs directly motivated by our observations.
Via extensive experiments on thirteen real-world hypergraphs (\cref{sec:experiments}), we validate the empirical superiority of the proposed method \ours.

\smallsection{Limitations and discussions.}
We use the same topological features used in existing works~\citep{choe2023classification,zheng2024co} for a fair comparison (see \cref{sec:method:stage1}), while finding more informative topological features is a potential future direction.

We assume no input node/hyperedge features, which is also true for the original datasets we use. However, when additional input node/hyperedge features are given, they can be directly incorporated in Stage 1 as additional features in $X$ and further improve the performance of \ours (see Sections~\ref{sec:method:stage1} and \ref{sec:experiments:acc}).

We mainly consider the scenarios where each group contains a single anchor since is it common in the real world (see Sections~\ref{sec:intro} and \ref{sec:problem_state}).
See Section~\ref{sec:experiments:acc} for discussions and results considering multiple anchors, where \ours still outperforms the baseline methods.
See Appendix~\ref{appx:discussions} for discussions on more general hypergraphs, e.g., directed hypergraphs and heterophilic hypergraphs.

\section*{Acknowledgments}
{\small This work was partly supported by the National Research Foundation of Korea (NRF) grant funded by the Korea government (MSIT) (No. RS-2024-00406985, 40\%). This work was partly supported by Institute of Information \& Communications Technology Planning \& Evaluation (IITP) grant funded by the Korea government (MSIT) (No. RS-2022-II220157, Robust, Fair, Extensible Data-Centric Continual Learning, 50\%) (No. RS-2019-II190075, Artificial Intelligence Graduate School Program (KAIST), 10\%).
The authors give thanks to Yijia Zheng (University of Amsterdam) for valuable discussions and for sharing the source code of CoNHD.
}

\normalem
\bibliographystyle{IEEEtran}
\bibliography{ref}

\clearpage
\onecolumn

\appendices

\crefname{section}{appendix}{appendices}%
\Crefname{section}{Appendix}{Appendices}%

\begin{center}
    {\large \textbf{Identifying Group Anchors in Real-World Group Interactions Under Label Scarcity: Appendix}}
\end{center}

{\small
\tableofcontents
}

\section{Additional Discussions on More General Hypergraphs (Supplementing Section~\ref{sec:prelim})}\label{appx:discussions}

Here, we provide additional discussions on more general hypergraphs, supplementing Section~\ref{sec:prelim}.

\smallsection{Directed hypergraphs.}
In directed hypergraphs, nodes within each hyperedge are partitioned into a source set and a destination set, and each hyperedge is considered to be from the nodes in the source set to the nodes in the destination set.
Specifically, directed hypergraphs can be used to model chemical reactions~\citep{jost2019hypergraph}, and the source set contains the reactant entities, which can be seen as entities that initialize the reaction, which is similar to the concept of group anchors in this work.

Specifically, for the real-world systems considered in this work:
\begin{itemize}[leftmargin=*]
    \item Co-authorship and movie cast datasets do not have intuitive directions;
    \item For online Q\&A datasets, the intuitive direction is from the questioner (which is the group anchor) to the answerers. Therefore, if we know the direction, we will immediately know the anchor in each group, and the problem of identifying group anchors will become trivial.
    \item Similarly, for email and social network datasets, the intuitive direction is from the sender (which is the group anchor) to the recipients, and the direction will immediately tell us the group anchors.
\end{itemize}

In general, the proposed algorithm \ours can be applied to directed graphs as follows:
\begin{itemize}[leftmargin=*]
    \item In Stage 1, when we calculate and collect topological features, directions can be considered. Specifically, for the topological features used in \ours:
    \begin{itemize}
        \item Node degrees can be divided into in-degrees and out-degrees;
        \item For eigenvector centrality and PageRank, we can consider the edge directions by using asymmetric adjacency matrices of directed graphs;
        \item For coreness, we can use the existing generalization of $k$-cores on directed graphs~\citep{giatsidis2013d};
    \end{itemize}
    \item In both Stage 1 and Stage 2, when we calculate and train anchor strengths, we can associate two (instead of one) anchor strengths to each node, representing its tendency to be the anchor when it is a head node and a tail node, respectively.  
\end{itemize}

\smallsection{Heterogeneous hypergraphs.}
Nodes or edges may have different classes/types in heterogeneous hypergraphs.
In such cases, there might be additional subtlety in defining group anchors.
We may define different group anchors for different classes/types.
Also, for some domains, classes/types may directly provide information on the significance of group members.

\begin{figure*}[h]
    \centering
    \includegraphics[width=0.8\linewidth]{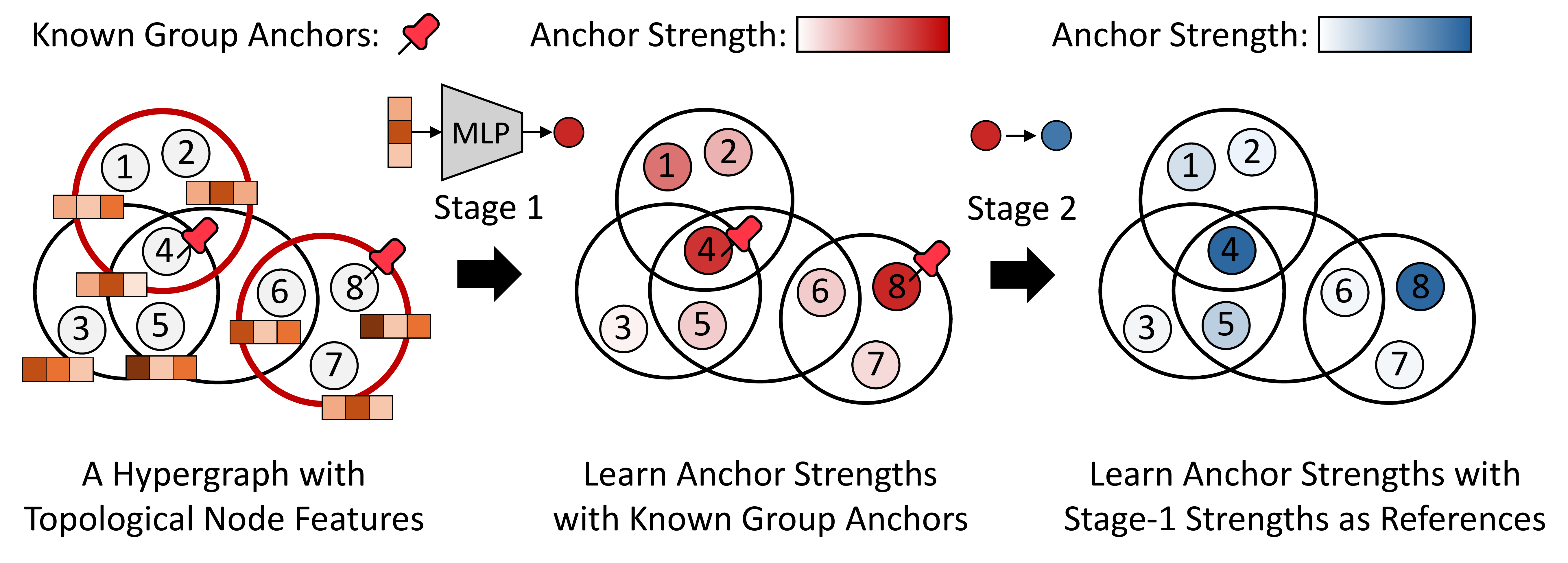}       
    \caption{An illustration of the proposed method \ours (Section~\ref{sec:method}).}
    \label{fig:method_illustration}    
\end{figure*}

\section{Additional Discussions on General Node Classification (Supplementing Section~\ref{sec:rel_wk})}\label{appx:general_node_cls}

Here, we provide more discussions on general node classification, complementing Section~\ref{sec:rel_wk}.

Node classification is a common task on both graphs~\citep{bhagat2011node} and hypergraphs~\citep{feng2019hypergraph}, where one aims to assign labels to nodes.
As deep learning techniques develop, the most popular methods for node classification on hypergraphs are hypergraph neural networks, including but not limited to 
HGNN~\citep{feng2019hypergraph}, 
HyperGCN~\citep{yadati2019hypergcn}, 
HNHN~\citep{dong2020hnhn}, 
HCHA~\citep{bai2021hypergraph}, and 
UniGNN~\citep{huang2021unignn}.
In general node classification, the node labels for a node are the same across different hyperedges, i.e., the node labels are edge-independent. The above existing methods for such general node classification cannot be directly applied to identifying group anchors, which are edge-dependent. For more details, please refer to recent surveys~\citep{antelmi2023survey,kim2024survey}.

\section{Additional Details on Group Roles and Group Anchors (Supplementing Section~\ref{sec:problem_state})}\label{appx:roles_and_seed_members}

Here, we provide additional details on group roles and group anchors, complementing Section~\ref{sec:problem_state}.

\smallsection{Domains.}
The real-world datasets used in our experiments are from five different domains:
\begin{enumerate}[leftmargin=*]
    \item Co-authorship domain: $D_{\text{co}}$;
    \item Online Q\&A domain: $D_{\text{qa}}$;
    \item Email domain: $D_{\text{em}}$;
    \item Social network domain: $D_{\text{so}}$;
    \item Movie cast domain: $D_{\text{mo}}$.
\end{enumerate}

\smallsection{Group roles.}
Below, for each domain $D$ considered in this work, we list the group roles $\calR(D)$ in the domain:
\begin{enumerate}[leftmargin=*]
    \item Co-authorship domain: $\calR(D_{\text{co}}) =$\;\{\texttt{firstAuthor},  \texttt{middleAuthor}, \texttt{lastAuthor}\};
    \item Online Q\&A domain: $\calR(D_{\text{qa}}) =$\;\{\texttt{questioner}, \texttt{answerer}\};
    \item Email domain: $\calR(D_{\text{em}}) =$\;\{\texttt{sender}, \texttt{TO-recipient}, \texttt{CC-recipient}\};
    \item Social network domain: $\calR(D_{\text{so}}) =$\;\{\texttt{initiator}, \texttt{participant}\}; for the Message (\texttt{soME}) dataset, for each group/hyperedge that represents a message, the initiator is the sender of the message, who initially chooses to include all the recipients; for the Retweet (\texttt{soRE}) dataset, for each group/hyperedge that represents a post, the initiator is the retweeted user, who uploads the original post so that the others can retweet it;
    \item Movie cast domain: $\calR(D_{\text{mo}}) =$\;\{\texttt{leadingActor}, \texttt{supportingActor}\}.
\end{enumerate}

\smallsection{Group anchors.}
The \textit{group-anchor function} $f_{\text{anchor}}$ maps each domain to the group role of anchors in that domain.
Given a domain $D \in \mathcal{D}$, $f_{\text{anchor}}(D) \in \mathcal{R}(D)$ is the group role of group anchors in domain $D$:
\begin{enumerate}
    \item $f_{\text{anchor}}(D_{\text{co}}) = \texttt{firstAuthor}~\text{or}~\texttt{lastAuthor}$;
    \item $f_{\text{anchor}}(D_{\text{qa}}) = \texttt{questioner}$;
    \item $f_{\text{anchor}}(D_{\text{em}}) = \texttt{sender}$;
    \item $f_{\text{anchor}}(D_{\text{so}}) = \texttt{initiator}$;
    \item $f_{\text{anchor}}(D_{\text{mo}}) = \texttt{leadingActor}$.
\end{enumerate} 

\begingroup
\setlength{\tabcolsep}{9pt}
\renewcommand{\arraystretch}{0.92}
\begin{table}[t!]
    \centering
    \caption{\uline{Observation 2: Whether a node is the group anchor or not is stable, yet not fully homogeneous, across different hyperedges.} The anchor purity in real-world datasets is significantly higher than in randomized ones in most cases, indicated by the small $p$-values obtained from one-sided $t$-tests, but the purity is not near 100\% in most cases.
    \revisekdd{Here, we report statistics on all hyperedge, supplementing the statistics in \cref{tab:label_purity}.}}
    \label{tab:obs_2_all}
\begin{adjustbox}{max width=\linewidth}
\begin{tabular}{cc|cc|c}
    \hline
    \multicolumn{2}{c|}{\textbf{Dataset}} & Real-world & Random & $p$-value \bigstrut\\
    \hline
    \multirow{2}[1]{*}{\texttt{coAA}} & (first) & 0.6913{\scriptsize$\pm$0.3482} & 0.5825{\scriptsize$\pm$0.3598} & $<$0.0001 \bigstrut[t]\\
          & (last) & 0.7306{\scriptsize$\pm$0.3392} & 0.5819{\scriptsize$\pm$0.3597} & $<$0.0001 \\
    \multirow{2}[0]{*}{\texttt{coDB}} & (first) & 0.6863{\scriptsize$\pm$0.3546} & 0.6043{\scriptsize$\pm$0.3555} & $<$0.0001 \\
          & (last) & 0.7501{\scriptsize$\pm$0.3307} & 0.6051{\scriptsize$\pm$0.3560} & $<$0.0001 \\
    \multirow{2}[0]{*}{\texttt{coSM}} & (first) & 0.7423{\scriptsize$\pm$0.3650} & 0.5728{\scriptsize$\pm$0.3987} & $<$0.0001 \\
          & (last) & 0.7753{\scriptsize$\pm$0.3515} & 0.5901{\scriptsize$\pm$0.4039} & $<$0.0001 \\
    \multicolumn{2}{c|}{\texttt{qaBI}} & 0.8241{\scriptsize$\pm$0.3050} & 0.5099{\scriptsize$\pm$0.3495} & $<$0.0001 \\
    \multicolumn{2}{c|}{\texttt{qaPH}} & 0.8216{\scriptsize$\pm$0.2929} & 0.5315{\scriptsize$\pm$0.3380} & $<$0.0001 \\
    \multicolumn{2}{c|}{\texttt{qaMA}} & 0.7700{\scriptsize$\pm$0.3720} & 0.6402{\scriptsize$\pm$0.3922} & 0.0076 \\
    \multicolumn{2}{c|}{\texttt{qaST}} & 0.8649{\scriptsize$\pm$0.3162} & 0.7029{\scriptsize$\pm$0.4021} & $<$0.0001 \\
    \multicolumn{2}{c|}{\texttt{emEN}} & 0.9209{\scriptsize$\pm$0.1836} & 0.8537{\scriptsize$\pm$0.2162} & $<$0.0001 \\
    \multicolumn{2}{c|}{\texttt{emEU}} & 0.6380{\scriptsize$\pm$0.1909} & 0.5499{\scriptsize$\pm$0.1372} & $<$0.0001 \\
    \multicolumn{2}{c|}{\texttt{emER}} & 0.7206{\scriptsize$\pm$0.1915} & 0.5352{\scriptsize$\pm$0.0935} & $<$0.0001 \\
    \multicolumn{2}{c|}{\texttt{soME}} & 0.7669{\scriptsize$\pm$0.3190} & 0.6728{\scriptsize$\pm$0.3473} & $<$0.0001 \\
    \multicolumn{2}{c|}{\texttt{soRE}} & 0.7444{\scriptsize$\pm$0.3533} & 0.5637{\scriptsize$\pm$0.3687} & $<$0.0001 \\
    \multicolumn{2}{c|}{\texttt{moML}} & 0.9925{\scriptsize$\pm$0.0738} & 0.4962{\scriptsize$\pm$0.3356} & $<$0.0001 \bigstrut[b]\\
    \hline
    \multicolumn{2}{c|}{Average} & 0.7775{\scriptsize$\pm$0.0875} & 0.5995{\scriptsize$\pm$0.0848} & - \bigstrut\\
    \hline
    \end{tabular}%
\end{adjustbox}
\end{table}
\endgroup

\section{Additional Details on Observation 2 (Supplementing Section~\ref{sec:intuition:global_consistency})}\label{appx:obs_2_all}

Here, we provide statistics on all hyperedges regarding Observation 2, supplementing Section~\ref{sec:intuition:global_consistency} (especially \cref{tab:label_purity}).

\section{Illustration of Proposed Method \ours (Supplementing Section~\ref{sec:method})}\label{appx:method_illustration}

Here, we provide an illustration of the proposed method \ours, supplementing Section~\ref{sec:method}.
See Figure~\ref{fig:method_illustration}.

\section{Additional Details on Feature Processing (Supplementing Section~\ref{sec:method:stage1})}\label{appx:feature_process}

Here, we provide additional details on feature processing in Stage 1 of \ours, supplementing Section~\ref{sec:method:stage1}.

Below are the details of feature normalization and aggregation used to obtain the features.
As mentioned in \cref{sec:method:stage1}, we use the same four topological features used by \citet{choe2023classification} and \citet{zheng2024co}:
(1) \textbf{node degree}, the number of hyperedges each node is in (see \cref{sec:prelim}),
(2 \& 3) \textbf{eigenvector centrality}~\citep{bonacich1987power} and \textbf{PageRank centrality}~\citep{page1999pagerank} computed on the weighted clique expansion, and (4) \textbf{coreness}~\citep{sun2020fully,seidman1983network}.
The feature processing on each raw topological feature is the same.
Let $x_{v}$ be the raw topological feature value of the node $v$.
For global features, the processed features for $v$ are the same across different hyperedges.
We normalize and/or aggregate each raw feature in six different ways:
\begin{itemize}[leftmargin=*]
    \item \textbf{(1) Global min-max normalization.} The processed feature 
    \[
    \tilde{x}^{(1)}_{v} = \frac{x_{v} - x_{min}}{x_{max} - x_{min}} \in [0, 1],
    \] 
    where
    $x_{max} = \max_{v' \in V} x_{v'}$ and
    $x_{min} = \min_{v' \in V} x_{v'}$.
    \item \textbf{(2) Global rank normalization.} We rank $x_{u}$'s for all the nodes $u \in V$ in zero-index and descending order, i.e., the largest value has rank $0$, the second-largest has rank $1$, etc.
    Let $r_{v}$ be the rank of $v$.
    The processed feature 
    \[
    \tilde{x}^{(2)}_{v} = \frac{r_{v}}{|V| - 1} \in [0, 1].
    \]    
    \item \textbf{(3\&4) Local min-max aggregated (mean and standard deviation).} First, we do min-max normalization in each hyperedge, i.e., $\hat{x}_{v;e} = \frac{x_{v;e} - x^{(e)}_{min}}{x^{(e)}_{max} - x^{(e)}_{min}}$ with
    $x^{(e)}_{max} = \max_{u \in e} x_{v;e}$ and
    $x^{(e)}_{min} = \min_{u \in e} x_{v;e}$.
    Then, we take the mean and standard deviation to obtain two aggregated features:
    \[
    \tilde{x}^{(3)}_{v} = \operatorname{Mean}(\hat{x}_{v;e'}: v \in e')
    \]
    and
    \[
    \tilde{x}^{(4)}_{v} = \operatorname{Stdev}(\hat{x}_{v;e'}: v \in e').
    \]    
    \item \textbf{(5\&6) Local rank aggregated (mean and standard deviation).}
    First, we compute normalized ranks in each hyperedge, i.e.,
    we rank $x_{u}$'s for all the nodes $u \in e$ in zero-index and descending order, let $\hat{r}_{v;e}$ be the rank of $v$ in $e$, and let $\hat{x}_{v;e} = \frac{\hat{r}_{v;e}}{|e| - 1}$.
    Then, we take the mean and standard deviation to obtain two aggregated features:
    \[
    \tilde{x}^{(5)}_{v} = \operatorname{Mean}(\hat{r}_{v;e'}: v \in e')
    \]
    and
    \[
    \tilde{x}^{(6)}_{v} = \operatorname{Stdev}(\hat{r}_{v;e'}: v \in e').
    \]    
\end{itemize}
For local features, the processed features might be distinct in different hyperedges, even for the same node.
We normalize and/or aggregate each raw feature in two different ways:
\begin{itemize}[leftmargin=*]
    \item \textbf{(1) Local min-max normalization.} We do min-max normalization in each hyperedge, i.e., 
    \[
    \tilde{x}^{(1)}_{v;e} = \frac{x_{v;e} - x^{(e)}_{min}}{x^{(e)}_{max} - x^{(e)}_{min}}
    \]
    with
    $x^{(e)}_{max} = \max_{u \in e} x_{v;e}$ and
    $x^{(e)}_{min} = \min_{u \in e} x_{v;e}$.    
    \item \textbf{(2) Local rank normalization.}
    We compute normalized ranks in each hyperedge, i.e.,    
    we rank $x_{u}$'s for all the nodes $u \in e$ in zero-index and descending order, let $\hat{r}_{v;e}$ be the rank of $v$ in $e$, and let
    \[
    \tilde{x}^{(2)}_{v;e} = \frac{\hat{r}_{v;e}}{|e| - 1}.
    \]    
\end{itemize}

\begingroup
\setlength{\tabcolsep}{5pt}
\begin{table*}[t]
    \centering    
    \caption{The fine-tuned hyperparameters of \ours.}
    \label{tab:hyperparams_ours}
\begin{adjustbox}{max width=\textwidth}
\begin{tabular}{c|cccccccccccccccc}
\hline
\multirow{2}[2]{*}{\textbf{Dataset}} & \multicolumn{2}{c}{\texttt{coAA}} & \multicolumn{2}{c}{\texttt{coDB}} & \multicolumn{2}{c}{\texttt{coSM}} & \multirow{2}[2]{*}{\texttt{qaBI}} & \multirow{2}[2]{*}{\texttt{qaPH}} & \multirow{2}[2]{*}{\texttt{qaMA}} & \multirow{2}[2]{*}{\texttt{qaST}} & \multirow{2}[2]{*}{\texttt{emEN}} & \multirow{2}[2]{*}{\texttt{emEU}} & \multirow{2}[2]{*}{\texttt{emER}} & \multirow{2}[2]{*}{\texttt{soME}} & \multirow{2}[2]{*}{\texttt{soRE}} & \multirow{2}[2]{*}{\texttt{moML}} \bigstrut[t]\\
      & (first) & (last) & (first) & (last) & (first) & (last) &       &       &       &       &       &       &       &       &       &  \bigstrut[b]\\
\hline
Stage 1 learning rate & 0.01  & 0.01  & 0.01  & 0.001 & 0.001 & 0.01  & 0.1   & 0.1   & 0.01  & 0.001 & 0.01  & 0.001 & 0.01  & 0.01  & 0.001 & 0.1 \bigstrut[t]\\
Stage 2 learning rate & 0.1   & 0.1   & 0.1   & 0.1   & 0.1   & 0.01  & 0.1   & 0.01  & 0.01  & 0.001 & 0.1   & 0.1   & 0.01  & 0.1   & 0.001 & 0.1 \\
$\alpha^{(2)}$ & 0.06  & 0.09  & 0.06  & 0.09  & 0.06  & 0.3   & 0.06  & 0.3   & 0.06  & 0.9   & 0.2   & 0.08  & 0.9   & 0.06  & 0     & 1 \\
$w^{(2)}$ & 7     & 4     & 10    & 3     & 6     & 2     & 4     & 4     & 1     & 9     & 9     & 2     & 3     & 10    & 1     & -* \bigstrut[b]\\
\hline
\multicolumn{8}{l}{\small *The best validation accuracy is obtained without global aggregation.} \\
\end{tabular}%
\end{adjustbox}
\end{table*}
\endgroup

\begingroup
\setlength{\tabcolsep}{6pt}
\begin{table*}[t]
    \centering
    \caption{The fine-tuned hyperparameters of WHATsNet.}
    \label{tab:hyperparams_whatsnet}
\begin{adjustbox}{max width=\textwidth}
\begin{tabular}{c|ccccccccccccc}
\hline
\textbf{Dataset} & \multicolumn{1}{l}{\texttt{coAA}} & \multicolumn{1}{l}{\texttt{coDB}} & \multicolumn{1}{l}{\texttt{coSM}} & \texttt{qaBI} & \texttt{qaPH} & \texttt{qaMA} & \texttt{qaST} & \texttt{emEN} & \texttt{emEU} & \texttt{emER} & \texttt{soME} & \texttt{soRE} & \texttt{moML} \bigstrut\\
\hline
Learning rate & 0.001 & 0.0001 & 0.0001 & 0.0001 & 0.0001 & 0.0001 & 0.001 & 0.001 & 0.001 & 0.001 & 0.001 & 0.001 & 0.0001 \bigstrut[t]\\
Batch size & 512   & 64    & 64    & 64    & 64    & 64    & 128   & 128   & 128   & 64     & 128     & 64     & 64 \\
Number of layers & 1     & 1     & 1     & 2     & 2     & 1     & 2     & 1     & 1     & 2    & 2   & 2    & 2 \bigstrut[b]\\
\hline
\end{tabular}%
\end{adjustbox}
\end{table*}
\endgroup

\begingroup
\setlength{\tabcolsep}{6pt}
\begin{table*}[t]
    \centering
    \caption{The fine-tuned hyperparameters of CoNHD-I.}
    \label{tab:hyperparams_conhd_i}
\begin{adjustbox}{max width=\textwidth}
\begin{tabular}{c|ccccccccccccc}
\hline
\textbf{Dataset} & \multicolumn{1}{l}{\texttt{coAA}} & \multicolumn{1}{l}{\texttt{coDB}} & \multicolumn{1}{l}{\texttt{coSM}} & \texttt{qaBI} & \texttt{qaPH} & \texttt{qaMA} & \texttt{qaST} & \texttt{emEN} & \texttt{emEU} & \texttt{emER} & \texttt{soME} & \texttt{soRE} & \texttt{moML} \bigstrut\\
\hline
Learning rate & 0.001 & 0.001 & 0.0001 & 0.0001 & 0.0001 & 0.001 & 0.0001 & 0.001 & 0.001 & 0.001 & 0.001 & 0.001 & 0.0001 \bigstrut[t]\\
Batch size & 256   & 128   & 64    & 64    & 64    & 64    & 128   & 64    & 128   & 64    & 128   & 128   & 64 \\
Number of layers & 1     & 1     & 2     & 2     & 1     & 2     & 2     & 1     & 1     & 2     & 2     & 2     & 2 \bigstrut[b]\\
\hline
\end{tabular}%
\end{adjustbox}
\end{table*}
\endgroup

\begingroup
\setlength{\tabcolsep}{6pt}
\begin{table*}[t]
    \centering
    \caption{The fine-tuned hyperparameters of CoNHD-U.}
    \label{tab:hyperparams_conhd_u}
\begin{adjustbox}{max width=\textwidth}
\begin{tabular}{c|ccccccccccccc}
\hline
\textbf{Dataset} & \multicolumn{1}{l}{\texttt{coAA}} & \multicolumn{1}{l}{\texttt{coDB}} & \multicolumn{1}{l}{\texttt{coSM}} & \texttt{qaBI} & \texttt{qaPH} & \texttt{qaMA} & \texttt{qaST} & \texttt{emEN} & \texttt{emEU} & \texttt{emER} & \texttt{soME} & \texttt{soRE} & \texttt{moML} \bigstrut\\
\hline
Learning rate & 0.001 & 0.001 & 0.0001 & 0.0001 & 0.0001 & 0.0001 & 0.0001 & 0.001 & 0.001 & 0.0001 & 0.001 & 0.001 & 0.001 \bigstrut[t]\\
Batch size & 256   & 128   & 64    & 64    & 64    & 64    & 64    & 128   & 128   & 128   & 128   & 64    & 128 \\
Number of layers & 1     & 2     & 2     & 2     & 1     & 1     & 2     & 1     & 1     & 2     & 2     & 1     & 2 \bigstrut[b]\\
\hline
\end{tabular}%
\end{adjustbox}
\end{table*}
\endgroup

\begingroup
\setlength{\tabcolsep}{6pt}
\begin{table*}[t]
    \centering
    \caption{The fine-tuned hyperparameters of HNHN.}
    \label{tab:hyperparams_hnhn}
\begin{adjustbox}{max width=\textwidth}
\begin{tabular}{c|ccccccccccccc}
\hline
\textbf{Dataset} & \multicolumn{1}{l}{\texttt{coAA}} & \multicolumn{1}{l}{\texttt{coDB}} & \multicolumn{1}{l}{\texttt{coSM}} & \texttt{qaBI} & \texttt{qaPH} & \texttt{qaMA} & \texttt{qaST} & \texttt{emEN} & \texttt{emEU} & \texttt{emER} & \texttt{soME} & \texttt{soRE} & \texttt{moML} \bigstrut\\
\hline
Learning rate & 0.001 & 0.001 & 0.001 & 0.001 & 0.001 & 0.0001 & 0.0001 & 0.001 & 0.001 & 0.0001 & 0.0001 & 0.0001 & 0.001 \bigstrut[t]\\
Batch size & 256   & 64    & 128   & 64    & 64    & 64    & 128   & 128   & 64    & 64    & 64    & 64    & 128 \\
Number of layers & 1     & 2     & 2     & 2     & 2     & 1     & 2     & 1     & 1     & 2     & 2     & 2     & 1 \bigstrut[b]\\
\hline
\end{tabular}%
\end{adjustbox}
\end{table*}
\endgroup

\begingroup
\setlength{\tabcolsep}{6pt}
\begin{table*}[t]
    \centering
    \caption{The fine-tuned hyperparameters of HGNN.}
    \label{tab:hyperparams_hgnn}
\begin{adjustbox}{max width=\textwidth}
\begin{tabular}{c|ccccccccccccc}
\hline
\textbf{Dataset} & \multicolumn{1}{l}{\texttt{coAA}} & \multicolumn{1}{l}{\texttt{coDB}} & \multicolumn{1}{l}{\texttt{coSM}} & \texttt{qaBI} & \texttt{qaPH} & \texttt{qaMA} & \texttt{qaST} & \texttt{emEN} & \texttt{emEU} & \texttt{emER} & \texttt{soME} & \texttt{soRE} & \texttt{moML} \bigstrut\\
\hline
Learning rate & 0.001 & 0.001 & 0.0001 & 0.001 & 0.001 & 0.0001 & 0.001 & 0.001 & 0.001 & 0.0001 & 0.001 & 0.001 & 0.001 \bigstrut[t]\\
Batch size & 256   & 64    & 128   & 128   & 64    & 64    & 128   & 64    & 64    & 128   & 64    & 64    & 64 \\
Number of layers & 2     & 2     & 2     & 1     & 2     & 2     & 1     & 1     & 1     & 2     & 1     & 1     & 1 \bigstrut[b]\\
\hline
\end{tabular}%
\end{adjustbox}
\end{table*}
\endgroup

\begingroup
\setlength{\tabcolsep}{6pt}
\begin{table*}[t]
    \centering
    \caption{The fine-tuned hyperparameters of HCHA.}
    \label{tab:hyperparams_hcha}
\begin{adjustbox}{max width=\textwidth}
\begin{tabular}{c|ccccccccccccc}
\hline
\textbf{Dataset} & \multicolumn{1}{l}{\texttt{coAA}} & \multicolumn{1}{l}{\texttt{coDB}} & \multicolumn{1}{l}{\texttt{coSM}} & \texttt{qaBI} & \texttt{qaPH} & \texttt{qaMA} & \texttt{qaST} & \texttt{emEN} & \texttt{emEU} & \texttt{emER} & \texttt{soME} & \texttt{soRE} & \texttt{moML} \bigstrut\\
\hline
Learning rate & 0.1   & 0.03  & 0.05  & 0.05  & 0.03  & 0.005 & 0.1   & 0.05  & 0.1   & 0.01  & 0.1   & 0.005 & 0.005 \bigstrut[t]\\
Dropout ratio & 0.3   & 0.7   & 0.5   & 0.7   & 0.7   & 0.3   & 0.3   & 0.7   & 0.7   & 0.3   & 0.3   & 0.7   & 0.5 \\
Number of layers & 1     & 2     & 2     & 2     & 2     & 2     & 1     & 1     & 1     & 1     & 2     & 1     & 1 \bigstrut[b]\\
\hline
\end{tabular}%
\end{adjustbox}
\end{table*}
\endgroup

\begingroup
\setlength{\tabcolsep}{6pt}
\begin{table*}[t]
    \centering
    \caption{The fine-tuned hyperparameters of HAT.}
    \label{tab:hyperparams_hat}
\begin{adjustbox}{max width=\textwidth}
\begin{tabular}{c|ccccccccccccc}
\hline
\textbf{Dataset} & \multicolumn{1}{l}{\texttt{coAA}} & \multicolumn{1}{l}{\texttt{coDB}} & \multicolumn{1}{l}{\texttt{coSM}} & \texttt{qaBI} & \texttt{qaPH} & \texttt{qaMA} & \texttt{qaST} & \texttt{emEN} & \texttt{emEU} & \texttt{emER} & \texttt{soME} & \texttt{soRE} & \texttt{moML} \bigstrut\\
\hline
Learning rate & 0.001 & 0.001 & 0.0001 & 0.0001 & 0.001 & 0.0001 & 0.001 & 0.001 & 0.001 & 0.001 & 0.001 & 0.001 & 0.001 \bigstrut[t]\\
Batch size & 512   & 64    & 128   & 64    & 64    & 128   & 128   & 128   & 64    & 128   & 128   & 64    & 64 \\
Number of layers & 1     & 2     & 2     & 1     & 1     & 2     & 2     & 1     & 1     & 2     & 2     & 2     & 1 \bigstrut[b]\\
\hline
\end{tabular}%
\end{adjustbox}
\end{table*}
\endgroup

\begingroup
\setlength{\tabcolsep}{6pt}
\begin{table*}[t]
    \centering
    \caption{The fine-tuned hyperparameters of UniGCN.}
    \label{tab:hyperparams_unigcn}
\begin{adjustbox}{max width=\textwidth}
\begin{tabular}{c|ccccccccccccc}
\hline
\textbf{Dataset} & \multicolumn{1}{l}{\texttt{coAA}} & \multicolumn{1}{l}{\texttt{coDB}} & \multicolumn{1}{l}{\texttt{coSM}} & \texttt{qaBI} & \texttt{qaPH} & \texttt{qaMA} & \texttt{qaST} & \texttt{emEN} & \texttt{emEU} & \texttt{emER} & \texttt{soME} & \texttt{soRE} & \texttt{moML} \bigstrut\\
\hline
Learning rate & 0.001 & 0.001 & 0.001 & 0.001 & 0.001 & 0.001 & 0.001 & 0.001 & 0.001 & 0.001 & 0.0001 & 0.0001 & 0.0001 \bigstrut[t]\\
Batch size & 512   & 64    & 64    & 128   & 64    & 64    & 128   & 128   & 128   & 128   & 64    & 64    & 128 \\
Number of layers & 2     & 2     & 1     & 2     & 2     & 2     & 2     & 1     & 1     & 2     & 1     & 2     & 1 \bigstrut[b]\\
\hline
\end{tabular}%
\end{adjustbox}
\end{table*}
\endgroup

\begingroup
\setlength{\tabcolsep}{6pt}
\begin{table*}[t]
    \centering
    \caption{The fine-tuned hyperparameters of HNN.}
    \label{tab:hyperparams_hnn}
\begin{adjustbox}{max width=\textwidth}
\begin{tabular}{c|ccccccccccccc}
\hline
\textbf{Dataset} & \multicolumn{1}{l}{\texttt{coAA}} & \multicolumn{1}{l}{\texttt{coDB}} & \multicolumn{1}{l}{\texttt{coSM}} & \texttt{qaBI} & \texttt{qaPH} & \texttt{qaMA} & \texttt{qaST} & \texttt{emEN} & \texttt{emEU} & \texttt{emER} & \texttt{soME} & \texttt{soRE} & \texttt{moML} \bigstrut\\
\hline
Learning rate & 0.03  & 0.05  & 0.005 & 0.05  & 0.01  & 0.05  & 0.03  & 0.03  & OOM*   & 0.005 & 0.005 & 0.1   & 0.05 \bigstrut[t]\\
Dropout ratio & 0.3   & 0.5   & 0.5   & 0.5   & 0.7   & 0.7   & 0.7   & 0.7   & OOM   & 0.3   & 0.5   & 0.7   & 0.7 \\
Number of layers & 1     & 1     & 2     & 1     & 1     & 1     & 1     & 1     & OOM   & 2     & 1     & 2     & 1 \bigstrut[b]\\
\hline
\multicolumn{8}{l}{\small *OOM = Out of memory.} \\
\end{tabular}%
\end{adjustbox}
\end{table*}
\endgroup

\begingroup
\setlength{\tabcolsep}{4pt}
\begin{table*}[t!]
    \centering
    \caption{\uline{\ours can achieve high accuracy even when trained and tested on different datasets.} The accuracy (\%) of \ours in inductive settings is highlighted in bold if it is better than that of the strongest baseline in transductive settings. Surprisingly, the performance of \ours in inductive settings is sometimes even better than in transductive settings.}
    \label{tab:inductive}
\begin{adjustbox}{max width=\linewidth}
\begin{tabular}{l|ccc|ccc|cc|cc|cc|cc}
\hline
\textbf{Training Dataset} & \multicolumn{3}{c|}{\texttt{qaBI}} & \multicolumn{3}{c|}{\texttt{qaPH}} & \multicolumn{2}{c|}{\texttt{coDB} (first)} & \multicolumn{2}{c|}{\texttt{coDB} (last)} & \multicolumn{2}{c|}{\texttt{coAA} (first)} & \multicolumn{2}{c}{\texttt{coAA} (last)} \bigstrut[t]\\
\textbf{Test Dataset} & \texttt{qaPH} & \texttt{qaMA} & \texttt{qaST} & \texttt{qaBI} & \texttt{qaMA} & \texttt{qaST} & \texttt{coAA} & \texttt{coSM} & \texttt{coAA} & \texttt{coSM} & \texttt{coDB} & \texttt{coSM} & \texttt{coDB} & \texttt{coSM} \bigstrut[b]\\
\hline
\ours (inductive) & 86.95 & 35.07 & 31.32 & \textbf{86.94} & \textbf{39.10} & 29.59 & \textbf{45.46} & \textbf{36.56} & \textbf{46.27} & \textbf{44.98} & 41.27 & 34.50 & 42.94 & 39.03 \bigstrut\\
\hline
\ours (transductive) & 88.93 & 41.16 & 35.12 & 87.23 & 41.16 & 35.12 & 49.77 & 38.33 & 49.32 & 44.91 & 46.02 & 38.33 & 50.72 & 44.91 \bigstrut[t]\\
Strongest baseline (transductive) & 88.07 & 38.99 & 31.98 & 85.35 & 38.99 & 31.98 & 44.74 & 35.66 & 46.10 & 40.87 & 42.92 & 35.66 & 45.83 & 40.87 \bigstrut[b]\\
\hline
\end{tabular}%
\end{adjustbox}
\end{table*}
\endgroup

\section{Additional Details on Complexity Analysis (Supplementing Section~\ref{sec:method:complexity})}\label{appx:complexity}

Here, we provide additional details on the complexity analysis of \ours, supplementing Section~\ref{sec:method:complexity}.
Below, we analyze the time and space complexities of \ours in detail.

\smallsection{Stage 1.}
We use a two-layer MLP with hidden dimension $D_h$.
The input topological feature matrix $X$ has a size of $\sum_{e \in E} |e| \times n_{f}$ with $n_{f}$ features.
A forward pass takes 
$O(\sum_{e \in E} |e| (n_{f} D_h + D_h + D_h + 1)) = O(\sum_{e \in E} |e| n_{f} D_h),$ 
and Eq.~\eqref{eq:loss_stage_1} takes
$O(\sum_{e \in E} |e|).$
Hence, the training (backward) time complexity is 
$O(n^{(1)}_{ep} \sum_{e \in E} |e| n_{f} D_h)$
and the inference (forward) takes
$O(\sum_{e \in E} |e| n_{f} D_h).$
The space complexity is
$O(\sum_{e \in E} |e| n_{f} + n_{f} D_h + D_h) = O(n_{f} (D_h + \sum_{e \in E} |e|)).$
The number of learnable parameters is 
$n_{f} D_h + D_h + D_h + 1 = D_h (n_{f} + 2) + 1.$

\smallsection{Stage 2.}
Eq.~\eqref{eq:loss_stage_2_label} takes 
$O(\sum_{e \in E} |e|)$.
Eq.~\eqref{eq:loss_stage_2_align} takes
$O(\sum_{e \in E} |e|)$.
Eq.~\eqref{eq:pred_member_prop} takes
$O(\sum_{e \in E} |e|)$.
Hence, the training (backward) time complexity is $O(n^{(2)}_{ep} \sum_{e \in E} |e|)$
and the inference (forward) takes
$O(\sum_{e \in E} |e|).$
The space complexity is
$O(|V| + \sum_{e \in E} |e|) = O(\sum_{e \in E} |e|).$
The number of learnable parameters is $|V|.$

\smallsection{Total.}
The training time complexity is
$O(\sum_{e \in E} (n^{(1)}_{ep} |e| n_{f} D_h + n^{(2)}_{ep})),$
the inference time complexity is
$O(\sum_{e \in E} |e| n_{f} D_h),$
the space complexity is
$O(n_{f} (D_h + \sum_{e \in E} |e|)),$ and
the number of learnable parameters is
$O(D_h (n_{f} + 2) + 1 + |V|).$

\smallsection{Notes.}
In our experiments, we use $D_h = 64$, $n_f = 33$ (see \cref{sec:method:stage1}), and $n^{(1)}_{ep} = n^{(2)}_{ep} = 100$.
For $|V|$ and $|e|$, see \cref{tab:dataset_summary}.

\section{Additional Details on Experimental Settings (Supplementing Section~\ref{sec:exp_settings})}\label{appx:exp_settings}

Here, we provide additional details on experimental settings, supplementing Section~\ref{sec:exp_settings}.

\smallsection{Hardware.}
All the experiments are run on a machine with 
two Intel Xeon\textsuperscript{\textregistered}   Silver 4214R (12 cores, 24 threads) processors,
a 512GB RAM,
and RTX 8000 D6 (48GB) GPUs.
A single GPU is used for all the experiments.

\smallsection{Data processing.}
Some datasets contain hyperedges of size one (i.e., a single node).
Such single-node groups are included when computing topological features but not included for training, validation, or testing.

\smallsection{Implementation.}
The implementation of the baseline methods is obtained from the official GitHub repositories of \citep{choe2023classification} and \citep{zheng2024co}: 
\url{https://github.com/young917/EdgeDependentNodeLabel} and \url{https://github.com/zhengyijia/CoNHD}.
All the methods are implemented in Python using Pytorch~\citep{paszke2019pytorch}.

\smallsection{Baseline methods.}
The baseline methods are trained on the original edge-dependent node classification (ENC) problem.
After training, each baseline method predicts a label distribution for each node-hyperedge pair $(v, e)$, i.e., 
\[
p(v, e, x^*) = \Pr[\text{$v$ has label $x$ in hyperedge $e$}]
\]
for different labels $x$.
Let $x_s$ be the label corresponding to group anchors.
We pick the node 
$v^* = \arg \max_v p(v, e, x^*)$ in each hyperedge $e$ as the predicted anchor.

\smallsection{Hyperparameter fine-tuning.}
We use validation data to fine-tune the hyperparameters of \ours and the baselines.
For \ours, we fix the number of iterations in both stages as $100$, and fine-tune the learning rates, the loss term coefficient $\alpha^{(2)}$, and the global aggregation weight $w^{(2)}$:
\begin{itemize}[leftmargin=*]
    \item Stage 1 learning rate: $\setbr{0.001, 0.01, 0.1}$;
    \item Stage 2 learning rate: $\setbr{0.001, 0.01, 0.1}$;
    \item Loss term coefficient $\alpha^{(2)}$: $\setbr{0, 0.01, 0.02, \ldots, 0.09, 0.1, 0.2, \ldots, 0.9, 1}$;
    \item Global aggregation weight $w^{(2)}$: $\setbr{1, 2, 3, \ldots, 10}$.
\end{itemize}
For the baselines, we follow the settings in original papers~\citep{choe2023classification,zheng2024co} for their fine-tuning.
For completeness, we repeat their fine-tuning settings below.
Specifically, for all the baselines, we fix the hidden dimension to $64$, the final embedding dimension to $128$, the number of the inducing points to $4$, the number of attention layers to $2$, and the dropout ratio to $0.7$.
We fine-tune their learning rate, batch size, and the number of layers.
Exceptionally, for HCHA~\cite{bai2021hypergraph} and HNN~\cite{aponte2022hypergraph}, we use full-batch training without sampling. We tune their hyperparameters in a larger search space for learning rates, the number of layers, and dropout ratios, while we fix the dimension of final node and hyperedge embeddings to $128$, the hidden dimension to $64$, and the number of epochs to $300$.
See Tables~\ref{tab:hyperparams_ours} to~\ref{tab:hyperparams_hnn} for the fine-tuned hyperparameters of all the methods.

\begingroup
\setlength{\tabcolsep}{4pt}
\renewcommand{\arraystretch}{0.92}
\begin{table*}[t!]
    \centering
    \caption{
    \uline{With a training ratio of 5\%, \ours still achieves higher accuracy in identifying group anchors than all the baseline in most cases.}
    For each setting, the best performance is highlighted in bold, while the second-best is underlined.
    The mean accuracy values (\%) over five random splits are reported with standard deviations. OOM represents ``out of memory''.}
    \label{tab:res_5p_train}
\begin{adjustbox}{max width=\textwidth}
    \begin{tabular}{c|cccccccccccccccc}
    \hline
    \multirow{2}[2]{*}{\textbf{Dataset}} & \multicolumn{2}{c}{\texttt{coAA}} & \multicolumn{2}{c}{\texttt{coDB}} & \multicolumn{2}{c}{\texttt{coSM}} & \multirow{2}[2]{*}{\texttt{qaBI}} & \multirow{2}[2]{*}{\texttt{qaPH}} & \multirow{2}[2]{*}{\texttt{qaMA}} & \multirow{2}[2]{*}{\texttt{qaST}} & \multirow{2}[2]{*}{\texttt{emEN}} & \multirow{2}[2]{*}{\texttt{emEU}} & \multirow{2}[2]{*}{\texttt{emER}} & \multirow{2}[2]{*}{\texttt{soME}} & \multirow{2}[2]{*}{\texttt{soRE}} & \multirow{2}[2]{*}{\texttt{moML}} \bigstrut[t]\\
          & (first) & (last) & (first) & (last) & (first) & (last) &       &       &       &       &       &       &       &       &       &  \bigstrut[b]\\
    \hline
    WHATsNet & 44.36 & 45.24 & \uline{42.11} & 45.45 & \uline{34.32} & 39.13 & \uline{85.05} & \uline{87.86} & 36.22 & 30.44 & \uline{49.71} & 50.87 & \uline{66.40} & \textbf{74.05} & \uline{97.55} & 40.91 \bigstrut[t]\\
    CoNHD-U & 44.82 & 45.00 & 40.54 & 42.59 & 30.91 & 39.63 & 78.39 & 76.68 & 29.99 & 26.62 & 44.56 & \textbf{52.12} & 65.03 & 73.33 & 96.36 & 41.73 \\
    CoNHD-I & \uline{44.89} & 44.81 & 40.17 & 42.15 & 30.91 & 39.63 & 78.54 & 76.78 & 34.83 & 25.09 & 44.05 & \uline{51.64} & 64.27 & 73.39 & 96.99 & \uline{42.01} \\
    HNHN  & 39.26 & 40.63 & 34.23 & 38.55 & 33.11 & 34.08 & 61.17 & 71.19 & 32.38 & 23.17 & 35.02 & 48.46 & 37.45 & 56.59 & 53.71 & 33.64 \\
    HGNN  & 44.68 & 44.64 & 41.51 & 44.19 & 33.02 & 37.78 & 78.15 & 81.10 & 30.49 & 30.26 & 33.52 & 49.26 & 39.55 & 62.08 & 82.91 & 36.73 \\
    HCHA  & 38.66 & 39.17 & 34.26 & 32.07 & 32.87 & 37.21 & 66.73 & 67.22 & 34.14 & 22.51 & 22.77 & 45.60 & 48.07 & 41.89 & 53.31 & 17.51 \\
    HAT   & 42.83 & \uline{45.26} & 35.31 & 36.87 & 30.81 & 32.91 & 74.67 & 80.89 & 27.69 & 22.41 & 47.26 & 50.44 & 44.64 & 66.39 & 90.40 & 34.90 \\
    UniGCN & 41.13 & 43.97 & 41.57 & \uline{45.63} & 31.60 & \uline{43.16} & 73.06 & 67.94 & \uline{39.30} & \uline{31.06} & 44.19 & 49.21 & 48.47 & 68.86 & 86.16 & 38.02 \\
    HNN   & 36.22 & 39.16 & 32.51 & 32.45 & 32.32 & 36.65 & 48.94 & 64.38 & 31.26 & 19.51 & 35.34 & OOM   & 45.34 & 52.62 & 63.71 & 26.00 \bigstrut[b]\\
    \hline
    \ours & \textbf{49.31} & \textbf{50.12} & \textbf{45.74} & \textbf{48.97} & \textbf{40.91} & \textbf{45.90} & \textbf{87.27} & \textbf{88.53} & \textbf{40.21} & \textbf{36.61} & \textbf{52.78} & 50.29 & \textbf{66.84} & \uline{73.85} & \textbf{97.66} & \textbf{42.91} \bigstrut\\
    \hline
    \end{tabular}%
\end{adjustbox}
\end{table*}
\endgroup

\begingroup
\setlength{\tabcolsep}{4pt}
\renewcommand{\arraystretch}{0.92}
\begin{table*}[t!]
    \centering
    \caption{
    \uline{With a training ratio of 2.5\%, \ours still achieves higher accuracy in identifying group anchors than all the baseline in most cases.}
    For each setting, the best performance is highlighted in bold, while the second-best is underlined.
    The mean accuracy values (\%) over five random splits are reported with standard deviations. OOM represents ``out of memory''.}
    \label{tab:res_2p5_train}
\begin{adjustbox}{max width=\textwidth}
    \begin{tabular}{c|cccccccccccccccc}
    \hline
    \multirow{2}[2]{*}{\textbf{Dataset}} & \multicolumn{2}{c}{\texttt{coAA}} & \multicolumn{2}{c}{\texttt{coDB}} & \multicolumn{2}{c}{\texttt{coSM}} & \multirow{2}[2]{*}{\texttt{qaBI}} & \multirow{2}[2]{*}{\texttt{qaPH}} & \multirow{2}[2]{*}{\texttt{qaMA}} & \multirow{2}[2]{*}{\texttt{qaST}} & \multirow{2}[2]{*}{\texttt{emEN}} & \multirow{2}[2]{*}{\texttt{emEU}} & \multirow{2}[2]{*}{\texttt{emER}} & \multirow{2}[2]{*}{\texttt{soME}} & \multirow{2}[2]{*}{\texttt{soRE}} & \multirow{2}[2]{*}{\texttt{moML}} \bigstrut[t]\\
          & (first) & (last) & (first) & (last) & (first) & (last) &       &       &       &       &       &       &       &       &       &  \bigstrut[b]\\
    \hline
    WHATsNet & \uline{45.72} & \uline{46.09} & \uline{41.72} & \uline{45.23} & \uline{34.57} & 37.43 & \uline{84.89} & \uline{86.92} & 33.74 & 29.83 & \uline{41.99} & 49.54 & \textbf{65.05} & \uline{72.36} & \textbf{97.17} & 40.79 \bigstrut[t]\\
    CoNHD-U & 42.43 & 43.09 & 39.17 & 40.79 & 31.57 & 30.08 & 76.30 & 73.72 & 23.35 & 21.57 & 38.18 & 49.60 & 50.21 & 67.09 & 94.51 & 40.18 \\
    CoNHD-I & 44.53 & 44.78 & 39.91 & 41.48 & 31.91 & 39.10 & 77.27 & 75.92 & 26.31 & 21.64 & 34.68 & \textbf{50.43} & \uline{64.36} & 71.47 & 94.39 & \uline{41.78} \\
    HNHN  & 38.90 & 40.46 & 34.70 & 36.91 & 31.25 & 38.17 & 60.40 & 67.30 & 24.73 & 21.74 & 32.78 & 46.98 & 46.24 & 52.87 & 53.17 & 33.26 \\
    HGNN  & 44.45 & 44.48 & 41.16 & 43.64 & 33.57 & 36.93 & 78.02 & 80.25 & 30.48 & 30.51 & 34.12 & 45.52 & 42.32 & 54.84 & 85.65 & 31.12 \\
    HCHA  & 37.65 & 38.49 & 33.33 & 32.24 & 33.64 & 35.61 & 59.72 & 52.13 & 28.54 & 23.49 & 16.90 & 46.05 & 43.68 & 41.89 & 52.98 & 20.62 \\
    HAT   & 41.09 & 44.39 & 33.49 & 34.91 & 30.64 & 32.86 & 58.60 & 78.65 & 28.03 & 22.54 & 41.18 & 49.24 & 43.45 & 65.67 & 83.86 & 32.28 \\
    UniGCN & 40.52 & 42.15 & 41.34 & 44.70 & 34.34 & \uline{40.50} & 75.97 & 74.95 & \uline{39.32} & \uline{30.79} & 40.10 & 47.27 & 48.36 & 66.70 & 82.33 & 38.52 \\
    HNN   & 36.31 & 37.61 & 34.05 & 33.40 & 33.67 & 34.68 & 40.10 & 53.23 & 31.07 & 21.06 & 27.93 & OOM   & 48.40 & 47.91 & 60.74 & 25.97 \bigstrut[b]\\
    \hline
    \ours & \textbf{48.93} & \textbf{49.42} & \textbf{44.41} & \textbf{47.92} & \textbf{41.18} & \textbf{45.35} & \textbf{86.90} & \textbf{88.25} & \textbf{39.62} & \textbf{36.15} & \textbf{47.53} & \uline{49.85} & 62.73 & \textbf{73.38} & \uline{96.28} & \textbf{41.87} \bigstrut\\
    \hline
    \end{tabular}%
\end{adjustbox}
\end{table*}
\endgroup

\section{Additional Results with Different Training Ratios (Supplementing Section~\ref{sec:experiments:acc})}\label{appx:diff_training_ratios}

Here, we provide additional experimental results with different training ratios, supplementing Section~\ref{sec:experiments:acc} (especially Table~\ref{tab:single_node_res}).
As mentioned in \cref{sec:exp_settings}, in the main results, the ratios of unique hyperedges for training, validation, and test are 7.5\%, 2.5\%, and 90\%, respectively.
We provide additional results with the following two settings:
\begin{itemize}[leftmargin=*]
    \item The ratios of unique hyperedges for training, validation, and test are \textbf{5\%}, 2.5\%, and \textbf{92.5\%}, respectively;
    \item The ratios of unique hyperedges for training, validation, and test are \textbf{2.5\%}, 2.5\%, and \textbf{95\%}, respectively.
\end{itemize}
The other experimental settings are kept the same.

See Table~\ref{tab:res_5p_train} for the results with a training ratio of 5\%, and see Table~\ref{tab:res_2p5_train} for the results with a training ratio of 2.5\%.
In both settings, \ours still outperforms the baseline methods in most cases.

\begingroup
\setlength{\tabcolsep}{4pt}
\renewcommand{\arraystretch}{0.92}
\begin{table*}[t!]
    \centering
    \vspace{-1mm}
    \caption{
    \uline{With NDCG as the evaluation metric, \ours still achieves better performance in identifying group anchors than all the baselines in most cases.}
    For each setting, the best performance is highlighted in bold, while the second-best is underlined.
    The mean NDCG values over five random splits are reported with standard deviations. OOM represents ``out of memory''.}
    \label{tab:res_ndcg}
\begin{adjustbox}{max width=\textwidth}
        \begin{tabular}{c|cccccccccccccccc}
    \hline
    \multirow{2}[2]{*}{\textbf{Dataset}} & \multicolumn{2}{c}{\texttt{coAA}} & \multicolumn{2}{c}{\texttt{coDB}} & \multicolumn{2}{c}{\texttt{coSM}} & \multirow{2}[2]{*}{\texttt{qaBI}} & \multirow{2}[2]{*}{\texttt{qaPH}} & \multirow{2}[2]{*}{\texttt{qaMA}} & \multirow{2}[2]{*}{\texttt{qaST}} & \multirow{2}[2]{*}{\texttt{emEN}} & \multirow{2}[2]{*}{\texttt{emEU}} & \multirow{2}[2]{*}{\texttt{emER}} & \multirow{2}[2]{*}{\texttt{soME}} & \multirow{2}[2]{*}{\texttt{soRE}} & \multirow{2}[2]{*}{\texttt{moML}} \bigstrut[t]\\
          & (first) & (last) & (first) & (last) & (first) & (last) &       &       &       &       &       &       &       &       &       &  \bigstrut[b]\\
    \hline
    WHATsNet & \uline{0.761} & 0.764 & \uline{0.742} & 0.753 & 0.703 & \uline{0.728} & \uline{0.944} & \textbf{0.953} & 0.699 & \uline{0.650} & \uline{0.783} & 0.808 & \uline{0.868} & \textbf{0.902} & 0.990 & 0.722 \bigstrut[t]\\
    CoNHD-U & 0.748 & 0.746 & 0.733 & 0.739 & 0.684 & 0.709 & 0.917 & 0.904 & 0.660 & 0.604 & 0.705 & \textbf{0.810} & 0.860 & 0.895 & 0.988 & 0.727 \\
    CoNHD-I & 0.756 & 0.758 & 0.734 & 0.743 & 0.674 & 0.719 & 0.920 & 0.910 & 0.665 & 0.611 & 0.710 & 0.808 & 0.857 & 0.896 & \uline{0.991} & \uline{0.729} \\
    HNHN  & 0.738 & 0.743 & 0.707 & 0.722 & 0.695 & 0.695 & 0.853 & 0.739 & 0.679 & 0.604 & 0.683 & 0.799 & 0.768 & 0.819 & 0.828 & 0.685 \\
    HGNN  & 0.759 & 0.765 & 0.738 & 0.750 & 0.700 & 0.722 & 0.928 & 0.900 & 0.680 & 0.645 & 0.717 & 0.799 & 0.764 & 0.839 & 0.943 & 0.700 \\
    HCHA  & 0.734 & 0.736 & 0.704 & 0.684 & 0.700 & 0.702 & 0.880 & 0.869 & 0.677 & 0.606 & 0.543 & 0.780 & 0.788 & 0.742 & 0.825 & 0.568 \\
    HAT   & 0.756 & \uline{0.765} & 0.720 & 0.730 & 0.683 & 0.695 & 0.905 & 0.925 & 0.664 & 0.606 & 0.776 & 0.808 & 0.767 & 0.869 & 0.972 & 0.688 \\
    UniGCN & 0.754 & 0.765 & 0.736 & \uline{0.755} & \uline{0.703} & 0.718 & 0.905 & 0.912 & \uline{0.707} & 0.648 & 0.751 & 0.800 & 0.824 & 0.869 & 0.956 & 0.711 \\
    HNN   & 0.728 & 0.736 & 0.687 & 0.711 & 0.697 & 0.714 & 0.852 & 0.840 & 0.663 & 0.613 & 0.717 & OOM   & 0.789 & 0.817 & 0.838 & 0.677 \bigstrut[b]\\
    \hline
    \ours & \textbf{0.782} & \textbf{0.786} & \textbf{0.758} & \textbf{0.774} & \textbf{0.730} & \textbf{0.770} & \textbf{0.949} & \uline{0.953} & \textbf{0.733} & \textbf{0.669} & \textbf{0.788} & \uline{0.809} & \textbf{0.875} & \uline{0.898} & \textbf{0.992} & \textbf{0.736} \bigstrut\\
    \hline
    \end{tabular}%
\end{adjustbox}
\end{table*}
\endgroup

\begingroup
\setlength{\tabcolsep}{4pt}
\renewcommand{\arraystretch}{0.92}
\begin{table*}[t!]
    \centering
    \vspace{-1mm}
    \caption{
    \uline{With MRR as the evaluation metric, \ours still achieves better performance in identifying group anchors than all the baselines in most cases.}
    For each setting, the best performance is highlighted in bold, while the second-best is underlined.
    The mean MRR values over five random splits are reported with standard deviations. OOM represents ``out of memory''.}
    \label{tab:res_mrr}
\begin{adjustbox}{max width=\textwidth}
    \begin{tabular}{c|cccccccccccccccc}
    \hline
    \multirow{2}[2]{*}{\textbf{Dataset}} & \multicolumn{2}{c}{\texttt{coAA}} & \multicolumn{2}{c}{\texttt{coDB}} & \multicolumn{2}{c}{\texttt{coSM}} & \multirow{2}[2]{*}{\texttt{qaBI}} & \multirow{2}[2]{*}{\texttt{qaPH}} & \multirow{2}[2]{*}{\texttt{qaMA}} & \multirow{2}[2]{*}{\texttt{qaST}} & \multirow{2}[2]{*}{\texttt{emEN}} & \multirow{2}[2]{*}{\texttt{emEU}} & \multirow{2}[2]{*}{\texttt{emER}} & \multirow{2}[2]{*}{\texttt{soME}} & \multirow{2}[2]{*}{\texttt{soRE}} & \multirow{2}[2]{*}{\texttt{moML}} \bigstrut[t]\\
          & (first) & (last) & (first) & (last) & (first) & (last) &       &       &       &       &       &       &       &       &       &  \bigstrut[b]\\
    \hline
    WHATsNet & \uline{0.680} & 0.685 & \uline{0.655} & 0.670 & 0.603 & \uline{0.637} & \uline{0.924} & \uline{0.936} & 0.600 & \uline{0.538} & \uline{0.710} & 0.741 & \uline{0.823} & \textbf{0.867} & 0.987 & 0.630 \bigstrut[t]\\
    CoNHD-U & 0.663 & 0.661 & 0.643 & 0.652 & 0.578 & 0.613 & 0.888 & 0.871 & 0.548 & 0.480 & 0.614 & \textbf{0.745} & 0.812 & 0.859 & 0.984 & 0.637 \\
    CoNHD-I & 0.674 & 0.677 & 0.645 & 0.656 & 0.565 & 0.625 & 0.892 & 0.878 & 0.555 & 0.489 & 0.621 & \uline{0.742} & 0.808 & 0.860 & \uline{0.988} & \uline{0.640} \\
    HNHN  & 0.649 & 0.656 & 0.609 & 0.629 & 0.593 & 0.593 & 0.802 & 0.649 & 0.572 & 0.478 & 0.582 & 0.729 & 0.687 & 0.757 & 0.767 & 0.582 \\
    HGNN  & 0.677 & 0.685 & 0.650 & 0.666 & 0.599 & 0.629 & 0.903 & 0.865 & 0.573 & 0.532 & 0.625 & 0.729 & 0.682 & 0.784 & 0.923 & 0.601 \\
    HCHA  & 0.644 & 0.646 & 0.605 & 0.579 & 0.599 & 0.603 & 0.839 & 0.823 & 0.571 & 0.481 & 0.407 & 0.705 & 0.714 & 0.653 & 0.763 & 0.428 \\
    HAT   & 0.673 & 0.685 & 0.626 & 0.639 & 0.576 & 0.593 & 0.871 & 0.899 & 0.553 & 0.482 & 0.701 & 0.741 & 0.687 & 0.824 & 0.962 & 0.586 \\
    UniGCN & 0.670 & \uline{0.685} & 0.647 & \uline{0.673} & \uline{0.608} & 0.632 & 0.874 & 0.882 & \uline{0.603} & 0.536 & 0.661 & 0.732 & 0.766 & 0.827 & 0.941 & 0.619 \\
    HNN   & 0.636 & 0.646 & 0.582 & 0.613 & 0.595 & 0.617 & 0.801 & 0.786 & 0.560 & 0.489 & 0.620 & OOM   & 0.716 & 0.754 & 0.786 & 0.572 \bigstrut[b]\\
    \hline
    \ours & \textbf{0.708} & \textbf{0.714} & \textbf{0.676} & \textbf{0.698} & \textbf{0.640} & \textbf{0.693} & \textbf{0.931} & \textbf{0.937} & \textbf{0.644} & \textbf{0.565} & \textbf{0.717} & 0.742 & \textbf{0.831} & \uline{0.863} & \textbf{0.989} & \textbf{0.649} \bigstrut\\
    \hline
    \end{tabular}%
\end{adjustbox}
\end{table*}
\endgroup

\section{Additional Results with Different Evaluation Metrics (Supplementing Section~\ref{sec:experiments:acc})}\label{appx:diff_eval_metrics}

Here, we provide additional experimental results with different evaluation metrics, supplementing Section~\ref{sec:experiments:acc} (especially Table~\ref{tab:single_node_res}).
Below, we provide results with two additional evaluation metrics: NDCG (normalized discounted cumulative gain) and MRR (mean reciprocal rank).

\smallsection{Ranking.}
To compute NDCG and MRR, we need to obtain a ranking of the nodes in each hyperedge. For \ours, the ranking is based on the anchor strength of each node. 
For the baselines, the ranking is based on the label logits of each node. 
Let the rankings be (1) descending, i.e., the node with the highest anchor strength or label logit is ranked first, and (2) 1-based, i.e., the rank of the first node is 1.

\smallsection{NDCG.}
The NDCG is calculated as the ratio of the DCG (discounted cumulative gain) to the IDCG (ideal DCG).
In each hyperedge $e = \{v_1, v_2, \ldots, v_k\}$, WLOG let $v_1$ be the ground-truth group anchor.
Then, for $e$, the relevance score of $v_1$ is $r_1 = 1$, and the relevance score of any other node is $r_i = 0, \forall i \neq 1$.
Let $a_i$ be the rank of $v_i$ in the ranking.
Then the DCG of $e$ is
$$\text{DCG}(e) = \sum_{i = 1}^{k} \frac{r_i}{\log_2(a_i+1)} = \frac{r_1}{\log_2(a_1+1)},$$
and the IDCG of $e$ is
$$\text{IDCG}(e) = \frac{1}{\log_2(1+1)} = 1.$$
Therefore, the NDCG of $e$ is
$$\text{NDCG}(e) = \frac{\text{DCG}(e)}{\text{IDCG}(e)} = \frac{r_1}{\log_2(a_1+1)}.$$
By averaging the NDCG of all hyperedges, we obtain the NDCG of the dataset.

\smallsection{MCC.}
The MRR is calculated as the reciprocal of the rank of the group anchor.
In each hyperedge $e = \{v_1, v_2, \ldots, v_k\}$, WLOG let $v_1$ be the ground-truth group anchor.
Let $a_i$ be the rank of $v_i$ in the ranking.
Then, the MRR of $e$ is
$$\text{MRR}(e) = \frac{1}{a_1}.$$
By averaging the MRR of all hyperedges, we obtain the MRR of the dataset.

See Table~\ref{tab:res_ndcg} for the results evaluated using NDCG, and see Table~\ref{tab:res_mrr} for the results evaluated using MRR.
In both settings, \ours still outperforms the baseline methods in most cases.

\begingroup
\setlength{\tabcolsep}{4pt}
\renewcommand{\arraystretch}{0.92}
\begin{table*}[t!]
    \centering
    \caption{
    \uline{The performance of \ours is stable and robust across different random seeds, with low standard deviations over five different random seeds.}
    For each random seed, the mean accuracy values (\%) over five random splits are reported.}
    \label{tab:res_random_seeds}
\begin{adjustbox}{max width=\textwidth}
    \begin{tabular}{c|cccccccccccccccc}
    \hline
    \multirow{2}[2]{*}{\textbf{Dataset}} & \multicolumn{2}{c}{\texttt{coAA}} & \multicolumn{2}{c}{\texttt{coDB}} & \multicolumn{2}{c}{\texttt{coSM}} & \multirow{2}[2]{*}{\texttt{qaBI}} & \multirow{2}[2]{*}{\texttt{qaPH}} & \multirow{2}[2]{*}{\texttt{qaMA}} & \multirow{2}[2]{*}{\texttt{qaST}} & \multirow{2}[2]{*}{\texttt{emEN}} & \multirow{2}[2]{*}{\texttt{emEU}} & \multirow{2}[2]{*}{\texttt{emER}} & \multirow{2}[2]{*}{\texttt{soME}} & \multirow{2}[2]{*}{\texttt{soRE}} & \multirow{2}[2]{*}{\texttt{moML}} \bigstrut[t]\\
          & (first) & (last) & (first) & (last) & (first) & (last) &       &       &       &       &       &       &       &       &       &  \bigstrut[b]\\
    \hline
    Seed 1 & 49.68 & 50.60 & 46.55 & 49.95 & 40.92 & 48.08 & 87.41 & 88.74 & 40.59 & 36.57 & 53.55 & 50.89 & 67.77 & 74.87 & 97.82 & 44.99 \bigstrut[t]\\
    Seed 2 & 49.65 & 50.58 & 46.66 & 49.96 & 40.86 & 48.11 & 87.34 & 88.84 & 40.67 & 36.60 & 53.96 & 50.81 & 66.70 & 74.97 & 97.82 & 43.36 \\
    Seed 3 & 49.64 & 50.58 & 46.64 & 49.96 & 40.99 & 48.30 & 87.37 & 88.67 & 40.03 & 36.60 & 52.57 & 50.85 & 66.50 & 74.81 & 97.82 & 43.38 \\
    Seed 4 & 49.67 & 50.57 & 46.67 & 49.97 & 40.44 & 48.19 & 87.42 & 88.66 & 38.39 & 36.75 & 54.14 & 51.02 & 66.49 & 74.95 & 97.82 & 43.35 \\
    Seed 5 & 49.72 & 50.58 & 46.59 & 49.96 & 41.19 & 47.93 & 87.36 & 88.81 & 39.71 & 36.74 & 53.11 & 50.90 & 68.33 & 74.86 & 97.82 & 43.38 \bigstrut[b]\\
    \hline
    Average & 49.67 & 50.58 & 46.62 & 49.96 & 40.88 & 48.12 & 87.38 & 88.74 & 39.88 & 36.65 & 53.47 & 50.89 & 67.16 & 74.89 & 97.82 & 43.69 \bigstrut[t]\\
    Stdev & 0.03  & 0.01  & 0.05  & 0.01  & 0.24  & 0.12  & 0.03  & 0.07  & 0.83  & 0.08  & 0.57  & 0.07  & 0.75  & 0.06  & 0.00  & 0.65 \bigstrut[b]\\
    \hline        
    \end{tabular}%
\end{adjustbox}
\end{table*}
\endgroup

\begin{figure*}[p]
    \centering
    \includegraphics[width=\linewidth]{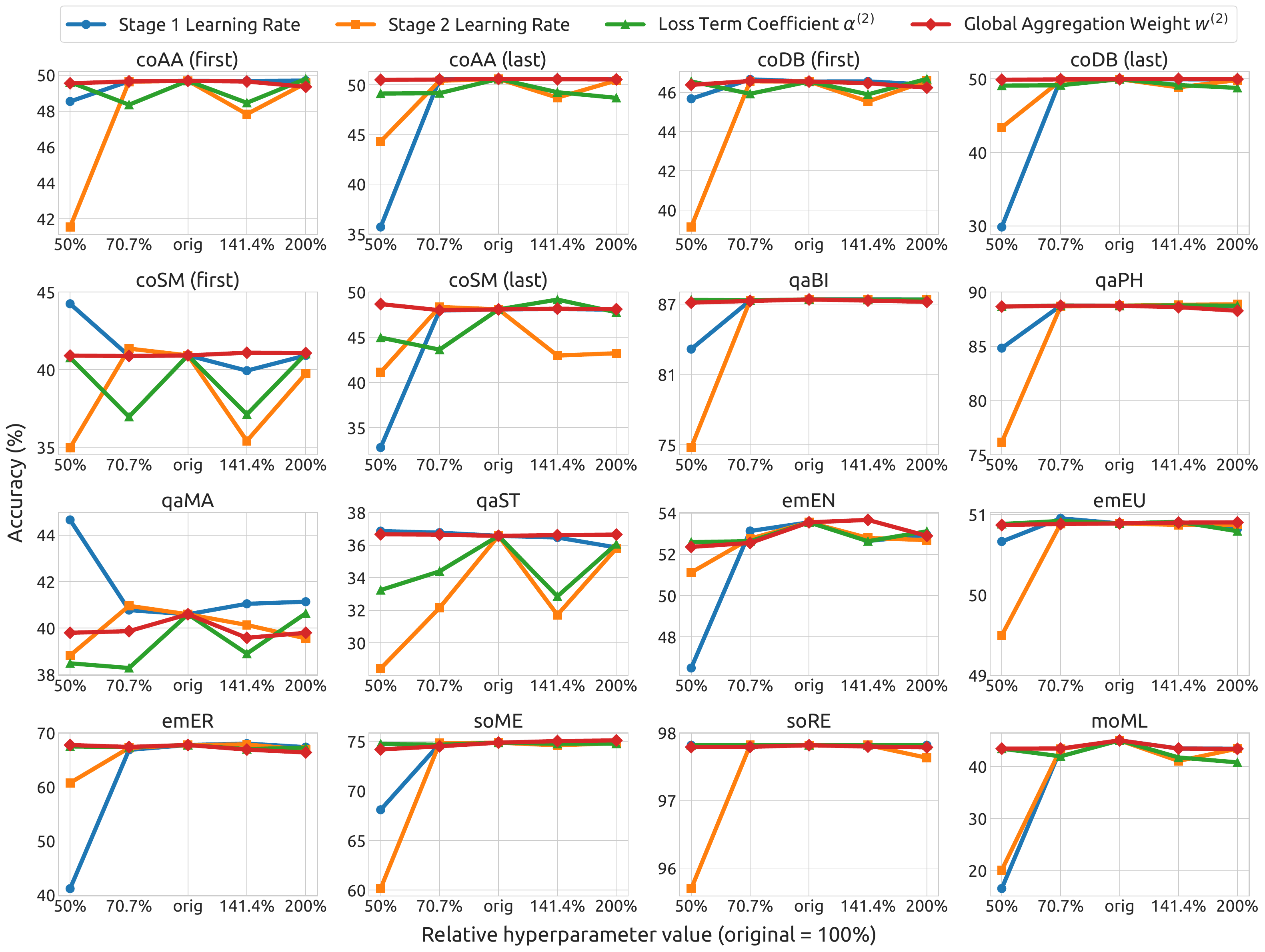}       
    \caption{Hyperparameter sensitivity of \ours. Each subplot corresponds to a dataset. For each hyperparameter, we plot the performance of \ours when manipulating the hyperparameter while keeping the other hyperparameters unchanged.}
    \label{fig:hyperparams}
\end{figure*}

\section{Additional Results with Multiple Random Seeds (Supplementing Section~\ref{sec:experiments:acc})}\label{appx:random_seeds}

Here, we provide additional experimental results with multiple random seeds, supplementing Section~\ref{sec:experiments:acc} (especially Table~\ref{tab:single_node_res}).
In Table~\ref{tab:single_node_res}, we show the performance of \ours with different training/validation/test splits and a single fixed random seed.
In Table~\ref{tab:res_random_seeds}, we provide the performance of \ours with different random seeds (still averaged over different training/validation/test splits).

\section{Additional Results on Hyperparameter Sensitivity (Supplementing Section~\ref{sec:experiments:acc})}\label{appx:hyperparam_sens}

Here, we provide additional experimental results on hyperparameter sensitivity, supplementing Section~\ref{sec:experiments:acc} (especially Table~\ref{tab:single_node_res}).
For each dataset, we examine the performance of \ours when we change each hyperparameter to $200$\%, $100\sqrt{2}$\% ($\approx 141.4$\%), $50\sqrt{2}$\% ($\approx 70.2$\%), and $50$\% of the originally fine-tuned value (see \cref{tab:hyperparams_ours}), while keeping the other hyperparameters unchanged.

See Figures~\ref{fig:hyperparams} for the changes in the performance when we change Stage 1 learning rate, Stage 2 learning rate, loss term coefficient $\alpha^{(2)}$, and global aggregation weight $w^{(2)}$.

\section{Additional Results on Inductive Settings (Supplementing Section~\ref{sec:experiments:acc})}\label{appx:experiments:acc}

Here, we provide additional experimental results on inductive settings, supplementing Section~\ref{sec:experiments:acc} (especially Table~\ref{tab:single_node_res}).

As mentioned in \cref{sec:experiments:acc}, we also consider inductive settings for \ours, i.e., \ours can be trained and tested on different datasets in the same domain.
We assume the same amount of known group anchors (i.e., 7.5\%) for the training dataset and no known group anchors for the test dataset.
We report the performance of the proposed method \ours in both transductive (i.e., trained and tested on the same dataset; the results in \cref{tab:single_node_res}) and inductive settings.
We use the performance of the strongest baseline in transductive settings as a reference.
As shown in \cref{tab:inductive}, even when trained and tested on different datasets (and using limited training data), it is possible for \ours to transfer knowledge on the correlations between topological features and group anchors.
Surprisingly, the performance of \ours in inductive settings is better than that of the strongest baseline in transductive settings in many cases and even better than the performance of \ours in transductive settings sometimes.

\begingroup
\setlength{\tabcolsep}{4pt}
\renewcommand{\arraystretch}{0.92}
\begin{table}[t!]
    \centering
    \caption{
    \uline{In scenarios with multiple anchors in each group, \ours consistently achieves higher accuracy in identifying group anchors than all the baseline.}
    For each setting, the best performance is highlighted in bold, while the second-best is underlined.
    The mean accuracy values (\%) over five random splits are reported with standard deviations.}
    \label{tab:res_multi}
\begin{adjustbox}{max width=\textwidth}
    \begin{tabular}{c|ccc}
    \hline
    \textbf{Dataset} & \texttt{coAA} & \texttt{coDB} & \texttt{coSM} \bigstrut\\
    \hline
    WHATsNet & 31.87 & 31.71 & 20.56 \bigstrut[t]\\
    CoNHD-U & \uline{34.03} & 32.97 & 19.57 \\
    CoNHD-I & 33.86 & \uline{33.06} & \uline{21.62} \\
    HNHN  & 26.88 & 25.72 & 21.01 \\
    HGNN  & 32.25 & 30.69 & 21.13 \\
    HCHA  & 25.58 & 24.10 & 23.30 \\
    HAT   & 29.57 & 24.25 & 20.02 \\
    UniGCN & 25.78 & 31.12 & 19.74 \\
    HNN   & 22.95 & 22.10 & 21.11 \bigstrut[b]\\
    \hline
    \ours & \textbf{37.50} & \textbf{37.42} & \textbf{32.81} \bigstrut\\
    \hline
    \end{tabular}%
\end{adjustbox}
\end{table}
\endgroup

\section{Additional Discussions and Results on Scenarios with Multiple Group Anchors (Supplementing Section~\ref{sec:intro} and Section~\ref{sec:experiments:acc})}\label{appx:multiple_seed_members}

Here, we provide additional discussions and experimental results on scenarios with multiple anchors in each group, supplementing Section~\ref{sec:intro} and Section~\ref{sec:experiments:acc}.

\smallsection{Experimental settings.}
To the best of our knowledge, no real-world group interaction datasets with multiple anchors in each group are available.
Therefore, we consider co-authorship datasets and set both the first and second authors as
anchors (motivated by scenarios with ``equal contribution'').
The accuracy is calculated as the proportion of groups where both anchors are correctly predicted.
See Table~\ref{tab:res_multi}.

\section{Additional Details on Downstream Application (Supplementing Section~\ref{sec:experiments:downstream})}\label{appx:downstream}

Here, we provide additional details on downstream application, supplementing Section~\ref{sec:experiments:downstream} (especially Table~\ref{tab:downstream}).

\smallsection{VilLain.}
For completeness, we include the details of VilLain~\citep{lee2024villain} below.
For more details, refer to the original paper~\citep{lee2024villain}.
The problem of hyperedge prediction is formulated as a binary classification task, predicting whether the given hyperedge is real or fake~\cite{patil2020negative,hwang2022ahp,yoon2020much}.
For each real hyperedge, VilLain generates a corresponding fake hyperedge with the same hyperedge size by randomly sampling subsets of nodes.
VilLain learns node embeddings.
To obtain the embedding of each hyperedge, VilLain applies maxmin pooling by (elementwise max pooling - elementwise min pooling) to the embeddings of the nodes in it.
After obtaining embeddings for both real and fake hyperedges, VilLain first trains a logistic regression classifier on training
hyperedges and their fake counterparts, and then tests on test hyperedges and their fake counterparts.
In each dataset, for group anchor identification, we have 7.5\% training hyperedges, 2.5\% validation hyperedges, and 90\% test hyperedges.
For this downstream application, we take 10\% hyperedges out of the test hyperedges as test hyperedges for group-interaction prediction.
VilLain uses the remaining hyperedges (7.5\% training, 2.5\% validation, and 80\% remaining test) to learn node embeddings.

\smallsection{Additional dimensions.}
As mentioned in \cref{sec:experiments:downstream}, based on the anchor strengths learned by \ours, we generate 11 additional dimensions, whose details we provide below.
Let $s^{(2)}_v$ denote the learned anchor strength of each node $v$.
First, we compute its within-edge normalized value, i.e.,
$\tilde{s}^{(2)}_{v;e} = \frac{\exp(s^{(2)}_v)}{\sum_{u \in e} s^{(2)}_u}$.
For each hyperedge $e$, we generate the following eleven additional dimensions:
\textbf{(1)} $\max_{v \in e} s^{(2)}_v$,
\textbf{(2)} $\min_{v \in e} s^{(2)}_v$,
\textbf{(3)} $\max_{v \in e} s^{(2)}_v - \min_{v \in e} s^{(2)}_v$,
\textbf{(4)} $\operatorname{Mean}(s^{(2)}_v: v \in e)$,
\textbf{(5)} $\operatorname{Stdev}(s^{(2)}_v: v \in e)$,
\textbf{(6)} $\sum_{v \in e} s^{(2)}_v$,
\textbf{(7)} $\max_{v \in e} \tilde{s}^{(2)}_{v;e}$,
\textbf{(8)} $\min_{v \in e} \tilde{s}^{(2)}_{v;e}$,
\textbf{(9)} $\max_{v \in e} \tilde{s}^{(2)}_{v;e} - \min_{v \in e} \tilde{s}^{(2)}_{v;e}$,
\textbf{(10)} $\operatorname{Mean}(s^{(2)}_v: v \in e)$,
and \textbf{(11)} $\operatorname{Stdev}(s^{(2)}_v: v \in e)$.

\end{document}